\documentclass[nofootinbib,preprint,tightenlines,superscriptaddress]{revtex4-1}

\usepackage[utf8]{inputenc}
\usepackage[english]{babel}
\usepackage{amssymb,amsthm,amsmath,amstext,amsbsy,amsopn}
\usepackage{bbm}
\usepackage{nicefrac}
\usepackage{slashed}
\usepackage{pstricks}
\usepackage{graphicx}
\usepackage{hyperref}
\usepackage{leftidx}
\usepackage{environ}
\usepackage{mathtools}
\usepackage{xspace}
\usepackage{overpic}

\newcommand{\ie}{\textit{i.e.}}
\newcommand{\eg}{\textit{e.g.}}
\newcommand{\cf}{\textit{cf.}\xspace}

\newcommand{\viz}{\textit{viz.}\xspace}
\newcommand{\apriori}{\textit{a priori}\xspace}

\newcommand{\mathspace}{\ \ }
\newcommand{\mathtext}[1]{\mathspace\text{#1}\mathspace}

\newcommand{\fm}{\ensuremath{\mathrm{fm}}}

\newcommand{\vecr}{\mathbf{r}}

\newcommand{\vecp}{\mathbf{p}}
\newcommand{\veck}{\mathbf{k}}
\newcommand{\vecq}{\mathbf{q}}

\newcommand{\vZero}{\mathbf{0}}

\newcommand{\dd}{\mathrm{d}}

\newcommand{\dq}[1]{\!\!\frac{\mathrm{d}^3#1}{(2\pi)^3}}

\newcommand{\ddq}{\dq{q}}

\newcommand{\gsim}{\gtrsim}

\newcommand{\ii}{\mathrm{i}}

\newcommand{\vD}{\boldsymbol{D}}
\newcommand{\hc}{\mathrm{h.c.}}
\newcommand{\OO}{\mathcal{O}}

\newcommand{\eps}{\varepsilon}

\newcommand{\EulerGamma}{C_E}

\newcommand{\Ip}{\mathrm{Im}}

\newcommand{\LiTwo}{\mathrm{Li}_2}

\newcommand{\MN}{M_N}

\newcommand{\Mpi}{M_\pi}

\newcommand{\yd}{y_d}
\newcommand{\yt}{y_t}

\newcommand{\sigmad}{\sigma_d}

\newcommand{\sigmadt}{\sigma_{d,t}}
\newcommand{\sigmatpp}{\sigma_{t,pp}}
\newcommand{\sigmatnn}{\sigma_{t,nn}}

\newcommand{\cd}{c_d}
\newcommand{\ct}{c_t}
\newcommand{\cdt}{c_{d,t}}
\newcommand{\ctpp}{c_{t,pp}}
\newcommand{\ctnn}{c_{t,nn}}

\newcommand{\gamd}{\gamma_d}

\newcommand{\rd}{\rho_d}

\newcommand{\app}{a_C}
\newcommand{\rpp}{r_C}

\newcommand{\ann}{\ensuremath{a_{t,nn}}}

\newcommand{\andDoublet}{\ensuremath{{}^2a_{\text{$n$--$d$}}}}

\newcommand{\ThreeSOne}{\ensuremath{{}^3S_1}\xspace}
\newcommand{\OneSNot}{\ensuremath{{}^1S_0}\xspace}

\newcommand{\Triton}{\ensuremath{{}^3\mathrm{H}}\xspace}
\newcommand{\ThreeH}{\Triton}
\newcommand{\ThreeHe}{\ensuremath{{}^3\mathrm{He}}\xspace}

\newcommand{\LamNoPi}{\Lambda_{\slashed\pi}}

\newcommand*\rvec[1]%
{\ensuremath{\overset{\smash{\raisebox{-1.5pt}{\tiny$\rightarrow$}}}{#1}}}
\newcommand*\lvec[1]%
{\ensuremath{\overset{\smash{\raisebox{-1.5pt}{\tiny$\leftarrow$}}}{#1}}}

\newcommand{\vk}{\mathbf{k}}
\newcommand{\vp}{\mathbf{p}}
\newcommand{\vq}{\mathbf{q}}

\newcommand{\Tgen}{\mathcal{T}}
\newcommand{\Bgen}{\mathcal{B}}
\newcommand{\Zgen}{\mathcal{Z}}
\newcommand{\Rgen}{\mathcal{R}}

\newcommand{\sss}{\mathrm{s}}
\newcommand{\ccc}{\mathrm{c}}
\newcommand{\fff}{\mathrm{full}}

\newcommand{\TS}{\Tgen_\sss}
\newcommand{\TSda}{\TS^\mathrm{d,a}}
\newcommand{\TSdb}{\TS^\mathrm{d,b1}}
\newcommand{\TSdc}{\TS^\mathrm{d,b2}}

\newcommand{\KS}{K_\sss}
\newcommand{\Kbub}{K_\text{bub}}
\newcommand{\KbubTwo}{K_\text{bub}^{(2)}}
\newcommand{\Kbox}{K_\text{box}}
\newcommand{\KboxTwo}{K_\text{box}^{(2)}}
\newcommand{\KtriIn}{K_\text{tri}^{(\text{in})}}
\newcommand{\KtriInTwo}{K_\text{tri}^{(\text{2,in})}}
\newcommand{\KtriOut}{K_\text{tri}^{(\text{out})}}
\newcommand{\KtriOutTwo}{K_\text{tri}^{(\text{2,out})}}
\newcommand{\Krd}{K_{\rho_d}}
\newcommand{\Krt}{K_{r_t}}

\newcommand{\TF}{\Tgen_\fff}
\newcommand{\TFda}{\TF^\mathrm{d,a}}

\newcommand{\vTS}{\vec{\mathcal{T}}_\sss}
\newcommand{\vTF}{\vec{\mathcal{T}}_\fff}

\newcommand{\MeV}{\ensuremath{\mathrm{MeV}}}
\newcommand{\GeV}{\ensuremath{\mathrm{GeV}}}

\newcommand{\atanh}{\mathrm{atanh}}

\newcommand{\diag}{\mathrm{diag}}

\newcommand{\LO}{\text{LO}\xspace}
\newcommand{\NLO}{\text{NLO}\xspace}
\newcommand{\NNLO}{\text{N$^2$LO}\xspace}
\newcommand{\NNNLO}{\text{N$^3$LO}\xspace}

\newcommand{\skrowspace}{0.5em}

\NewEnviron{subalign}[1][]{%
\begin{subequations}\begin{align}
  \BODY
\end{align}\label{#1}\end{subequations}
}

\NewEnviron{spliteq}{%
\begin{equation}\begin{split}
  \BODY
\end{split}\end{equation}
}

\newcommand*{\vcenteredhbox}[1]
{\begingroup\setbox0=\hbox{#1}\parbox{\wd0}{\box0}\endgroup}

\begin{document}

\title{Second-order perturbation theory for \ThreeHe and $pd$ scattering\\
in pionless EFT}

\author{Sebastian König}
\email{koenig.389@osu.edu}
\affiliation{Department of Physics, The Ohio State University,
Columbus, Ohio 43210, USA}
\affiliation{Institut für Kernphysik, Technische Universität Darmstadt, 
64289 Darmstadt, Germany}
\affiliation{ExtreMe Matter Institute EMMI,
GSI Helmholtzzentrum für Schwerionenforschung GmbH,
64291 Darmstadt, Germany}

\date{\today}

\begin{abstract}
This work implements pionless effective field theory with the two-nucleon 
system expanded around the unitarity limit at second-order perturbation theory.
The expansion is found to converge well.  All Coulomb effects are treated in 
perturbation theory, including two-photon contributions at 
next-to-next-to-leading order.  After fixing a three-nucleon force to the 
\ThreeHe binding energy at this order, proton-deuteron scattering in the 
doublet S-wave channel is calculated for moderate center-of-mass momenta.
\end{abstract}

\maketitle

\section{Introduction}
\label{sec:Intro}

Effective field theory (EFT) is an established tool in theoretical nuclear 
physics to connect the description of nuclei in terms of hadronic degrees of 
freedom to the underlying physics of quantum chromodynamics (QCD).  In relying 
only on basic concepts---symmetries, the separation of scales, and a systematic 
ordering of contributions---it is both elegant and pragmatic at the same 
time.  While the most widely used framework used for the description of nuclear 
structure and reactions is based on the approximate (and spontaneously 
broken) chiral symmetry of the QCD, for momentum scales well below the 
pion mass $\Mpi$, pion-exchange contributions cannot be resolved explicitly.  
The resulting pionless EFT contains, besides electromagnetic forces, only 
contact interactions between nonrelativistic
fields~\cite{Bedaque:1997qi,vanKolck:1997ut,Kaplan:1998tg,Bedaque:1998mb,
Kaplan:1998we,Birse:1998dk,vanKolck:1998bw,Chen:1999tn,Bedaque:1999vb,
Gabbiani:1999yv}.  From the original chiral symmetry it only keeps the isospin 
subgroup as an approximate symmetry.

This, however, is not the primary driving feature of the theory.  Instead, the 
fact that the $NN$ S-wave scattering lengths---$5.4$ (${-}23.7$)~\fm\ in 
the \ThreeSOne (\OneSNot) channel---happen to be large compared to the typical 
scale $\Mpi^{-1} \sim 1.4~\fm$ implies that the few-nucleon sector is close to 
a universal regime where short-range details have little impact on low-energy 
parameters.  In the two-nucleon sector, this is manifest in the effective range 
expansion (ERE) working as well as it does~\cite{Bethe:1949yr}.  In the 
three-nucleon sector, the spin-doublet S-wave configuration is governed 
by a non-derivative three-body interaction at leading order
(\LO)~\cite{Bedaque:1999ve,Hammer:2000nf,Hammer:2001gh,Bedaque:2002yg,
Afnan:2003bs,Griesshammer:2005ga}, and the theory, formally equivalent to 
one for three bosons with short-range interactions, describes the triton 
as an approximate Efimov state~\cite{Efimov:1970zz,Efimov:1981aa,Hammer:2010kp}. 
For a recent review of the pionless three-body calculations, see 
Ref.~\cite{Vanasse:2016jtc}.

An important feature of the theory is that it can be constructed in a way such 
that it is fully renormalized order by order, with all 
nonperturbative effects included at \LO and the rest treated in perturbation 
theory~\cite{Hammer:2001gh,Vanasse:2013sda,Konig:2013cia,Vanasse:2014kxa,
Konig:2015aka}.  This allows for precise calculations of low-energy 
observables with controlled error estimates based directly on the EFT expansion 
and independent of working at a fixed or limited regulator scale.  Recent 
examples of high-order calculations can be found in 
Refs.~\cite{Margaryan:2015rzg,Vanasse:2015fph}).  Moreover, pionless EFT 
provides a controlled ``laboratory'' to study such perturbative schemes 
(expanding directly the amplitude instead of a potential), which have 
been argued to be important for chiral EFT as well (see for example 
Refs.~\cite{Long:2016ybt,Valderrama:2016koj} for recent reviews).
  
While electromagnetic effects are certainly important for a realistic 
description of nuclei, their inclusion into the EFT provides a challenge 
because the long-range nature of these forces---meaning that the Coulomb force 
becomes dominant at very low energies, precisely where the short-range 
expansion otherwise works best---is not easily accommodated in the power 
counting.  Early work on Coulomb effects in the 
$pp$~\cite{Kong:1998sx,Kong:1999sf,Kong:2000px,Butler:2001jj,Barford:2002je,
Ando:2007fh,Ando:2008va,Ando:2008jb} and $pd$~\cite{Rupak:2001ci} systems was 
followed up on in Refs.~\cite{Ando:2010wq,Konig:2011yq}, with the latter 
providing a first calculation of $pd$ doublet-channel scattering.

It has become clear that if perturbative renormalization is to be maintained, 
including electromagnetic effects is not as simple as adding a Coulomb 
potential to the short-range terms, as it is typically done in calculations 
based on effective pionless potentials~\cite{Kirscher:2009aj,Kirscher:2011uc,
Kirscher:2011zn,Lensky:2016djr}.  Based on studying the regulator (cutoff) 
dependence of the amplitude, it was realized that in the presence of 
nonperturbative Coulomb effects an isospin breaking three-nucleon force is 
required to ensure renormalization at next-to-leading order
(\NLO)~\cite{Vanasse:2014kxa,Konig:2014ufa}.  While another 
calculation~\cite{Kirscher:2015zoa} does not see the need to include such a 
term, Ref.~\cite{Konig:2015aka} showed that in part this three-body force is 
related to a two-body divergence in the $pp$ sector, and that the theory can be 
rearranged in such a way that all Coulomb effects in the \ThreeHe bound state 
are included in perturbation theory.  In this counting the \OneSNot channel is 
taken in the unitarity limit (infinite scattering length) at \LO and the effects 
of the finite physical scattering lengths---$a_t$ as given above as well as the 
Coulomb-modified $pp$ scattering $a_C \simeq 7.81~\fm$---are accounted for by 
parameters entering at \NLO.  In particular, in the $pp$ channel this parameter 
absorbs the logarithmic divergence generated by one-photon exchange in the 
two-nucleon subsystem.  Demoting all other (small) isospin-breaking effects to 
next-to-next-to-leading order (\NNLO) or higher, Ref.~\cite{Konig:2015aka} was 
able to calculate the \ThreeH--\ThreeHe binding-energy difference at \NLO 
without a new three-nucleon force.  More generally, this scheme enhances the 
predictive power of the theory by making the \OneSNot channel parameter free at 
\LO.

More recent work~\cite{Konig:2016utl} goes further and considers a more 
radical expansion that takes the \ThreeSOne scattering length to infinity as well 
in a ``full unitarity'' leading order, whereas Ref.~\cite{Vanasse:2016umz} 
explores similar ideas by expanding around a leading order that exhibits the 
$SU(4)_W$ spin-isospin symmetry~\cite{Wigner:1936dx} for finite scattering 
lengths.

In this work, the pionless unitarity-\LO counting schemes developed in 
Refs.~\cite{Konig:2015aka,Konig:2016utl} are implemented up to \NNLO, including 
effects from two-photon exchange and other isospin-breaking corrections.  It 
thereby establishes the convergence of these expansions up to this order.  
Going beyond \NLO (first-order perturbation theory), where all corrections 
enter linearly, is an important proof of principle.  Moreover, it is 
demonstrated explicitly that away from the zero-energy threshold it is possible 
to describe $pd$ scattering with fully perturbative Coulomb effects.  Finally, 
it establishes the presence of an \NNLO three-body force, to be fixed by a 
single $pd$ doublet datum, at \NNLO.

\medskip
After discussing the basic setup and contributions---most notably two-photon 
diagrams---in Sec.~\ref{sec:Formalism}, the perturbative calculation of 
three-body observables is described in Sec.~\ref{sec:ThreeBody}.  Results
are presented and discussed in Sec.~\ref{sec:Results}, followed by a conclusion 
and outlook in Sec.~\ref{sec:Conclusion}.  Some details left out from the main 
text are provided in an appendix.

\section{Formalism and building blocks}
\label{sec:Formalism}

This section collects the ingredients required to set up the \NNLO 
calculation, summarizing results from previous works as far as necessary to 
make the current description reasonably self-contained.

\subsection{Effective Lagrangian and dibaryon propagators}

Using the notation of Ref.~\cite{Konig:2015aka}, the effective Lagrangian is 
split into one-, two- and three-body terms according to
\begin{equation}
 \mathcal{L} = N^\dagger\left(\ii D_0+\frac{\vD^2}{2\MN}\right)N
 +\mathcal{L}_\mathrm{2d}+\mathcal{L}_\mathrm{2t}+\mathcal{L}_3 \,,
\label{eq:L-Nd}
\end{equation}
where $N$ is the nonrelativistic nucleon field (doublet in spin and isospin 
space), coupled to (Coulomb) photons with the covariant derivative $D_\mu = 
\partial_\mu + \ii eA_\mu \hat{Q}_N$ (with charge operator 
$\hat{Q}_N=(1+\tau_3)/2$).  The $NN$ S-wave two-body interactions are 
conveniently expressed in terms of auxiliary dibaryon fields $d^i$ (\ThreeSOne) 
and $t^A$ (\OneSNot), where $i$ and $A$ are spin-1 and isospin-1 indices, 
respectively.  The corresponding Lagrangians are
\begin{equation}
 \mathcal{L}_\mathrm{2d} = -d^{i\dagger}\left[\sigmad
 + \cd\left(\ii D_0+\frac{\vD^2}{4\MN}\right)\right]d^i
 + \yd\left[d^{i\dagger}\left(N^T P^i_d N\right)+\hc\right] \,,
\label{eq:L-3S1}
\end{equation}
with the repeated superscripts $i$ summed over, and
\begin{multline}
 \mathcal{L}_\mathrm{2t} = -t^{0\dagger}\left[\sigma_t
 + \ct\left(\ii D_0+\frac{\vD^2}{4\MN}\right)\right]t^0
 - t^{{-1}\dagger}\left[\sigmatpp
 + \ctpp\left(\ii D_0+\frac{\vD^2}{4\MN}\right)\right]t^{-1}\\
 - t^{{+1}\dagger}\left[\sigmatnn
 + \ctnn\left(\ii D_0+\frac{\vD^2}{4\MN}\right)\right]t^{+1}
 +\yt\left[t^{{\tilde A}\dagger}
 \left(N^T \tilde{P}^{\tilde A}_t N\right)+\hc\right] \,.
\label{eq:L-1S0}
\end{multline}
Note that the \OneSNot part is separated into physical channels ($np$, $nn$, 
$pp$).  Further details, including the projection operators $P^i_d$ and 
$\tilde{P}^{\tilde{A}}_t$ can be found in Ref.~\cite{Konig:2015aka}.  Setting
$\yd^2 = \yt^2 = 4\pi/\MN$, the remaining low-energy constants 
$\sigma_{d/t(,\cdot\cdot)}$ and $c_{d/t(,\cdot\cdot)}$ correspond directly to 
the scattering lengths and effective ranges in the respective channel.  
Contributions from the $NN$ shape parameters first enter at 
\NNNLO~\cite{Margaryan:2015rzg} in the standard pionless power counting so that 
they also do not have to be included in the calculations presented here.

In the well-known fashion, nucleon-bubble insertions into the bare dibaryon 
propagators are resummed to obtain the full leading-order expressions
\begin{equation}
 \ii\Delta_{d/t(,\cdot\cdot)}^{(0)}(p_0,\vp)
 = \frac{-\ii}{\sigma_{d/t(,\cdot\cdot)}^{(0)}
 + y_{d/t(,\cdot\cdot)}^2 I_0(p_0,\vp)} \,,
\label{eq:Delta-dt}
\end{equation}
where
\begin{multline}
 I_0(p_0,\vp) = \MN\int^\Lambda\ddq
 \frac{1}{\MN p_0 - \vp^2/4 - \vq^2 + \ii\eps} \\
 = {-}\frac{\MN}{4\pi}
 \left(\frac{2\Lambda}{\pi} - \sqrt{\frac{\vp^2}{4}-\MN p_0-\ii\eps}\right)
 +\OO(1/\Lambda)
\label{eq:I0-cutoff}
\end{multline}
is the generic nucleon bubble integral (Green's function from zero to zero 
separation) calculated with a sharp momentum cutoff.

In principle, an S-D mixing operator in the \ThreeSOne channel enters at \NNLO. 
However, at this order it does not contribute to the \ThreeH--\ThreeHe binding 
energy splitting, and neither is it relevant for the doublet S-wave 
phase shift~\cite{Vanasse:2013sda}.  Hence, it is not necessary to consider 
this operator in the present work.  Moreover, terms stemming from relativistic 
corrections and transverse photons are suppressed by inverse powers of $\MN$ 
and do thus not contribute to the order considered in this paper.

\subsubsection{Power counting around the unitarity limit}

In the power counting developed in Ref.~\cite{Konig:2015aka}, which is applied 
here up to \NNLO, the usual pionless expansion in terms of $Q/\LamNoPi$,
where $Q$ denotes the typical momentum of the process under 
consideration and $\LamNoPi \sim \Mpi$ is the pionless breakdown scale,
is paired with an additional expansion in $\aleph_0/Q$, where
\begin{equation}
 \aleph_0 \sim \alpha\MN \sim 1/|a_t| \sim 1/|\app|
\end{equation}
is a typical low-energy scale in the \OneSNot channel.  This combines the 
inverse scattering lengths with the typical Coulomb scale $\alpha\MN$.  In 
particular, at leading order the \OneSNot channel is considered in the 
unitarity limit (where the scattering lengths are infinite) and thus also 
isospin symmetric.  The expansion in $\aleph_0/Q$ then corresponds to including 
the effects of finite $1/a_t$ and $1/\app$ in perturbation theory, where the 
latter is naturally paired with one-photon exchange and ensures consistent 
renormalization in the presence of Coulomb effects.  Range corrections reflect 
the $Q/\LamNoPi$ expansion, as in the standard pionless counting.  Details of 
how the expansion is implemented by fixing the dibaryon parameters are given in 
Sec.~\ref{sec:TwoBodyParamaters}.

This scheme is constructed for the regime $\aleph_0 \ll Q \ll \LamNoPi$, which 
naturally includes the \ThreeHe bound states.  This was studied in 
Ref.~\cite{Konig:2015aka}, which also found that the $\aleph_0/Q$ expansion
works well in the $nd$ scattering system.  In $pd$ scattering, Coulomb effects 
are certainly nonperturbative at very small center-of-mass momenta ($k\to0$), 
but for larger $k$ (determining $Q$ in this case), the perturbative expansion 
should work as well.  This is demonstrated in the present work.

More generally, the scheme of Ref.~\cite{Konig:2015aka} counts isospin-breaking 
corrections, not only of electromagnetic origin but also those induced by the 
up-down quark-mass difference in QCD.  As required by renormalization, the 
effects that give rise $a_C \neq a_t$ to are accounted for at \NLO together 
with electromagnetic contributions.  Different values have been determined for 
the $nn$ scattering length~\cite{GonzalezTrotter:2006wz,Huhn:2001yk}, but it is 
generally assumed large (and negative), such that $1/\ann \approx 1/a_t$.  
Following the counting of Ref.~\cite{Konig:2015aka}, this difference is 
accounted for here by an \NNLO correction, and for definiteness the central 
value favored by the pionless analysis of Ref.~\cite{Kirscher:2011zn}, 
$\ann \simeq {-}(22.9\pm 4.1)~\fm$, is adopted here.  The small isospin 
breaking in the effective ranges, $r_C - r_t \simeq 0.06~\fm$ is also included 
at \NNLO.\footnote{The individual values used here are 
$r_C=2.794~\fm$~\cite{Bergervoet:1988zz} and 
$r_t=2.73~\fm$~\cite{Preston:1975}.}

\medskip
Recently, Ref.~\cite{Konig:2016utl} suggested a more radical rearrangement 
of the power counting that includes the \ThreeSOne inverse scattering length 
$1/a_d$ in $\aleph_0$ instead of counting it as $\OO(Q)$ (which is done in the 
standard pionless counting).  Although this expansion only perturbatively 
moves the deuteron bound state---which at \LO is located at zero energy in this 
expansion---to its physical position, Ref.~\cite{Konig:2016utl} found that 
it works well for three and four-nucleon bound states at \NLO.  As part of 
this work, \ThreeHe is considered up to second order in this ``full unitarity'' 
scheme.

\subsubsection{Three-nucleon forces}

It is well known that in pionless EFT a three-nucleon interaction is needed 
already at \LO to ensure renormalization of the doublet-channel 
amplitude~\cite{Bedaque:1999ve}.  This piece can be written
as~\cite{Ando:2010wq,Griesshammer:2011md}
\begin{equation}
 \mathcal{L}_{3,\LO} = \frac{h}{3}N^\dagger\left[\yd^2\,
 d^{i\dagger} d^j \sigma^i \sigma^j+\yt^2\,t^{A\dagger} t^B \tau^A\tau^B
 - \yd\yt\left(d^{i\dagger} t^A \sigma^i \tau^A + \hc\right) \right]N \,,
\label{eq:L-3}
\end{equation}
where the coupling $h$ is also split up in various orders, with each piece 
having a characteristic log-periodic dependence on the momentum cutoff 
$\Lambda$ that is used in the three-body integral equations:
\begin{equation}
 h = \frac{2H_{0,0}(\Lambda)}{\Lambda^2}
 + \frac{2H_{0,1}(\Lambda)}{\Lambda^2}
 + \frac{2H_{0,2}(\Lambda)}{\Lambda^2}
 + \cdots \,.
\end{equation}
In addition to this, Ref.~\cite{Ji:2012nj} firmly established (for the 
analogous three-boson system) that an additional energy-dependent three-body 
force enters at \NNLO.  Following Ref.~\cite{Vanasse:2013sda}, where this was 
worked out for the $nd$ doublet system, it is included here in the form
\begin{equation}
 h_2 = \frac{2H_{2,2}(\Lambda)}{\Lambda^4}(\MN E + \gamd^2) \,,
\end{equation}
which conveniently vanishes at the $Nd$ scattering threshold and thus 
simplifies the numerical parameter determination.  These standard three-body
force are not changed by expanding around the unitarity limit or the 
perturbative inclusion of Coulomb effects.

\medskip
While the \NLO isospin-breaking three-nucleon force identified 
in~\cite{Vanasse:2014kxa} is not needed in the counting scheme employed here 
(due to the fully perturbative treatment of Coulomb effects and only using 
isospin-symmetric effective ranges in the \OneSNot channel at that order), it 
turns out (see Sec.~\ref{sec:ThreeBodyEquations}) that such a term is 
eventually needed at \NNLO.  In Sec.~\ref{sec:Results} it is shown that a term 
of the form
\begin{equation}
 h_\alpha = \frac{2H_{0,2}^{(\alpha)}(\Lambda)}{\Lambda^2} \,,
\end{equation}
with an operator structure designed to act only in the $pd$ doublet channel but 
otherwise give the same factors as for $h$~\cite{Vanasse:2014kxa}, is 
sufficient to ensure renormalization at \NNLO.

\subsection{Second-order binding energy shifts}
\label{sec:SecondOrderEnergies}

The off-shell amplitude, determined by the Lippmann--Schwinger equation 
describing nucleon-deuteron scattering (see Sec.~\ref{sec:ThreeBodyEquations}), 
is the central object considered in this paper.  It describes both scattering 
(in the on-shell limit) as well as bound-state properties.  The binding energy 
of a given state can be obtained from it by expanding the expression that 
formally includes all orders, as discussed in Ref.~\cite{Ji:2012nj}.  In the 
limit where the energy $E$ approaches the bound-state pole, it can be written as
(with $k$ and $p$ denoting in- and outgoing momenta, respectively)
\begin{multline}
 \Tgen^{(0)}(E;k,p) + \Tgen^{(1)}(E;k,p) + \Tgen^{(2)}(E;k,p) + \cdots \\
 = {-}\frac{\Zgen^{(0)}(k,p) + \Zgen^{(1)}(k,p) + \Zgen^{(2)}(k,p) + \cdots}
 {E + B_0 + B_1 + B_2 + \cdots}
 + \Rgen^{(0)}(E;k,p) + \cdots \,.
\label{eq:T-pole-012}
\end{multline}
In this expression, $\Zgen^{(n)}$ and $\Rgen^{(n)}$, respectively, denote 
residue and regular terms, and the superscripts label contributions from 
different orders (note that regular terms with $n>0$ have been omitted in 
Eq.~\eqref{eq:T-pole-012}).  Expanding this by formally factoring out a small 
parameter and matching orders naturally recovers
\begin{equation}
 \Tgen^{(0)}(E;k,p) = {-}\frac{\Zgen^{(0)}(k,p)}{E + B_0} + \Rgen^{(0)}(E;k,p)
 \mathtext{with} \Zgen^{(0)}(k,p) = \Bgen^{(0)}(k)\,\Bgen^{(0)}(p)
\label{eq:T0-B0}
\end{equation}
at leading order.  Using the vertex functions $\Bgen^{(0)}$, which can be 
obtained directly by solving a homogeneous integral equation and properly 
normalizing the solutions~\cite{Konig:2011yq}, one can write down the various 
matrix elements that contribute to the \NLO energy shift.  This corresponds 
directly to first-order perturbation theory with bound-state wavefunctions and 
is what was done in Refs.~\cite{Konig:2014ufa,Konig:2015aka}.
Ref.~\cite{Ji:2012nj} instead starts from the \NLO matching condition,
\begin{multline}
 \Tgen^{(1)}(E;k,p)
 = B_1 \frac{\Zgen^{(0)}(k,p)}{(E + B_0)^2} + \frac{\Zgen^{(1)}(k,p)}{E + B_0}
 + \Rgen^{(1)}(E;k,p) \\
 \implies B_1 = \lim\nolimits_{E\to{-}B_0} \frac{(E+B_0)^2\,\Tgen^{(1)}(E;k,p)}
 {\Bgen^{(0)}(k)\,\Bgen^{(0)}(p)} \,,
\label{eq:B1-lim}
\end{multline}
and simplifies this to recover again the explicit expression in terms of 
$\Bgen^{(0)}$.  At \NNLO,
\begin{equation}
 B_2 = \lim\nolimits_{E\to{-}B_0}
 \frac{(E+B_0)^2\,\Tgen^{(2)}(E;k,p) + B_1(E+B_0)\Tgen^{(1)}(E;k,p)}
 {\Bgen^{(0)}(k)\,\Bgen^{(0)}(p)} \,,
\label{eq:B2-lim}
\end{equation}
which Ref.~\cite{Ji:2012nj} expands explicitly in terms of $\Bgen^{(0)}$ and 
$\Rgen^{(0)}(E;k,p)$.  Since the expressions become rather involved at this 
point (particularly so with the inclusion of Coulomb corrections for \ThreeHe), 
it is more convenient to directly evaluate Eq.~\eqref{eq:B2-lim}.  Numerical 
solutions at any given order can be obtained very efficiently using the 
technique described in Ref.~\cite{Vanasse:2013sda}.  In 
Ref.~\cite{Konig:2013cia}, it was extended to include Coulomb contributions and 
applied to calculate the quartet-channel $pd$ scattering up to \NNLO.

\subsection{Two-body T-matrix at second order}
\label{sec:TwoBody-NNLO}

To demonstrate how this procedure works in general, it is instructive to 
take as an example the \ThreeSOne nucleon-nucleon system expanded around 
the unitarity limit, which was considered recently in Ref.~\cite{Konig:2016utl}. 
The calculation here also serves to prove explicitly what was claimed for the 
\NNLO deuteron energy shift in that paper.

Switching temporarily to a pionless formalism without dibaryon fields, the 
momentum-independent contact interaction $C_0$ is expanded as
\begin{equation}
 C_0 = C_0^{(0)} + C_0^{(1)} +  C_0^{(2)} + \cdots \,.
\label{eq:C0-expansion}
\end{equation}
No explicit label is included here to keep the notation simple because this 
subsection considers only a single channel.

%%%%%%%%%%%%%%%%%%%%%%%%%%%%%%%%%%%%%%%%%%%%%%%%%%%%%%%%%%%%%%%%%%%%%%%%%%%%%%
\begin{figure}[htbp]
\centering
\begin{minipage}{0.95\textwidth}
\includegraphics[height=3.3em]{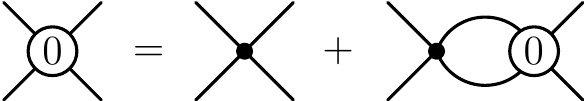}\\[0.66em]
\includegraphics[height=3.3em]{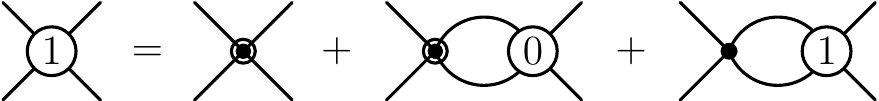}\\[0.66em]
\includegraphics[height=3.3em]{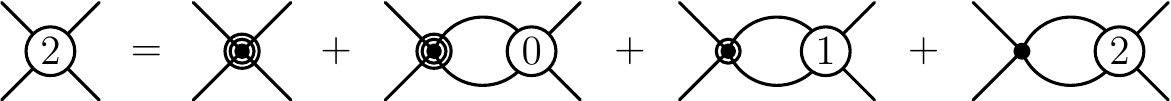}
\end{minipage}
\caption{Diagrammatic $NN$ integral equations up to second order.  Solid 
lines represent nucleons and an open circles with number $n$ denotes the $n$-th 
order T-matrix part.  A dot with $i$ surrounding circles corrections to a 
$C_0^{(i)}$ vertex.}
\label{fig:NN-IntEq-012}
\end{figure}
%%%%%%%%%%%%%%%%%%%%%%%%%%%%%%%%%%%%%%%%%%%%%%%%%%%%%%%%%%%%%%%%%%%%%%%%%%%%%%

Furthermore ignoring range corrections (\ie, momentum-dependent contact terms) 
for simplicity, the integral equations that define the T-matrix up to second 
order in the approach of Ref.~\cite{Vanasse:2013sda} are shown in 
Fig.~\ref{fig:NN-IntEq-012}.  Note that lower-order amplitudes appear as part of 
the \NLO and \NNLO inhomogeneous terms, but that the integral-equation 
kernel---determined solely by $C_0^{(0)}$---is always the same.  Iterating 
these equations shows that they recover the standard distorted-wave amplitudes 
that can also be obtained by a direct calculation of the individual diagrams at 
each order.  The primary advantage of the integral-equation formalism is that 
it avoids the explicit calculation of a two-loop integral (involving a full 
off-shell \LO T-matrix), which was the primary motivation for its introduction 
in Ref.~\cite{Vanasse:2013sda}.\footnote{In the two-body sector with separable 
regularization, even full off-shell contributions become trivial and can be 
calculated analytically.  In the three-body sector, however, using the 
integral-equation technique is a substantial simplification.}

All two-nucleon bubbles in Fig.~\ref{fig:NN-IntEq-012} correspond to the 
integral $I_0$ given in Eq.~\eqref{eq:I0-cutoff}.  With the sharp cutoff 
regulator (or more generally any separable one), the integral equations at each 
order can be solved algebraically.  At leading order (first line in 
Fig.~\ref{fig:NN-IntEq-012}), one simply recovers the well-known result
\begin{equation}
 \Tgen^{(0)}(E) = {-}C_0^{(0)} + C_0^{(0)} I_0(E)\,\Tgen^{(0)}(E)
 = \frac{{-}C_0^{(0)}}{1 - C_0^{(0)}I_0(E)} \,,
\label{eq:T-NN-0}
\end{equation}
corresponding directly to the propagator expression~\eqref{eq:Delta-dt}.  At 
\NLO and \NNLO, the equations and solutions are
\begin{equation}
 \Tgen^{(1)}(z) = {-}C_0^{(1)} + C_0^{(1)} I_0(z)\,\Tgen^{(0)}(z)
 + C_0^{(0)} I_0(z)\,\Tgen^{(1)}(z) \\
 = \frac{{-}C_0^{(1)}\left[1 + I_0(z)\,\Tgen^{(0)}(z)\right]}
 {1 - C_0^{(0)}I_0(z)}
\label{eq:T-NN-1}
\end{equation}
and
\begin{multline}
 \Tgen^{(2)}(z) = {-}C_0^{(2)} + C_0^{(2)} I_0(z)\,\Tgen^{(0)}(z)
 + C_0^{(1)} I_0(z)\,\Tgen^{(1)}(z)
 + C_0^{(0)} I_0(z)\,\Tgen^{(2)}(z) \\
 = \frac{{-}C_0^{(2)}\left[1 + I_0(z)\,\Tgen^{(0)}(z)\right]
 - C_0^{(1)}I_0(z)\,\Tgen^{(1)}(z)}{1 - C_0^{(0)}I_0(z)} \,.
\label{eq:tau-2}
\end{multline}

The leading-order term $C_0^{(0)}$, chosen to get a deuteron 
state with binding momentum $\kappa = \kappa^{(0)} \to 0$, corresponds to 
approaching the unitarity limit from the side of positive scattering length 
(\ie, a bound state moves towards zero energy).  At \NLO, $C_0^{(1)}$ is 
adjusted to get the first term in the effective range expansion given by the 
physical scattering length, $\kappa^{(1)} = 1/a_d$.  These conditions are 
satisfied by setting
\begin{equation}
 C_0^{(0)} = {-}\frac{2\pi^2}{\MN\Lambda}
 \left(1 - \frac{\pi\kappa^{(0)}}{2\Lambda}\right)^{-1} \mathtext{,}
 C_0^{(1)} = {-}\frac{\MN\kappa^{(1)}}{4\pi} \big(C_0^{(0)}\big)^2 \,.
\label{eq:C00-C01}
\end{equation}
At \NNLO, they are maintained with
\begin{equation}
 C_0^{(2)}
 = \left(\frac{\MN\kappa^{(1)}}{4\pi}\right)^2 \big(C_0^{(0)}\big)^3 \,.
\label{eq:C02}
\end{equation}
Using a different regularization scheme (\eg, dimensional regularization with 
power divergence subtraction~\cite{Kaplan:1998tg} or the Gaussian regulator 
used for the four-body calculations in Ref.~\cite{Konig:2016utl}) only changes 
the details of $C_0^{(0)}$ but not the general features of the calculation.  
By construction, this gives
\begin{equation}
 \Tgen^{(0)}(E)
 = \frac{4\pi}{\MN\left(\sqrt{\mathstrut{-}\MN E}-\kappa^{(0)}\right)}
 = \frac{8\pi\kappa^{(0)}}{\MN^2\left(E + \frac{(\kappa^{(0)})^2}{\MN}\right)}
 + \text{reg.\ terms}
\label{eq:T-0}
\end{equation}
and
\begin{equation}
 \Tgen^{(1)}(E) = \frac{4\pi\kappa^{(1)}}
 {\MN\left(\sqrt{\mathstrut{-}\MN E}-\kappa^{(0)}\right)^{2}}
 \mathtext{,}
 \Tgen^{(2)}(E) = \frac{4\pi\left(\kappa^{(1)}\right)^2}
 {\MN\left(\sqrt{\mathstrut{-}\MN E}-\kappa^{(0)}\right)^{3}} \,.
\label{eq:T-12}
\end{equation}
The \LO binding energy and the shifts up to second order are found to be
\begin{equation}
 B_0 = \frac{(\kappa^{(0)})^2}{\MN}
 \mathtext{,}
 B_1 = \frac{2\kappa^{(0)}\kappa^{(1)}}{\MN}
 \mathtext{,}
 B_2 = \frac{(\kappa^{(1)})^2}{\MN}
\end{equation}
from Eqs.~\eqref{eq:T0-B0}, \eqref{eq:B1-lim}, and~\eqref{eq:B2-lim}.  In 
particular, in the unitarity limit $\kappa^{(0)}\to0$ the deuteron remains at 
zero energy at \NLO, but it can be seen explicitly that it moves to its 
zero-range position, $B_2 \sim (\kappa^{(1)})^2 = 1/a_d^2$, at \NNLO,
exactly as claimed in Ref.~\cite{Konig:2016utl}.  While the results in 
Eqs.~\eqref{eq:T-0} and~\eqref{eq:T-12} match what one would naïvely expect 
from a direct expansion of the renormalized amplitude (\ie, written in terms of 
$\kappa = \kappa^{(0)} + \kappa^{(1)}$), it is reassuring to see the 
binding-energy shifts come out as desired even though in the limit 
$\kappa^{(0)}\to0$ the deuteron disappears as a state with normalizable 
wavefunction.

\subsection{Two-photon contributions}

\begin{figure}[htbp]
\centering
\includegraphics[width=0.85\textwidth,clip]{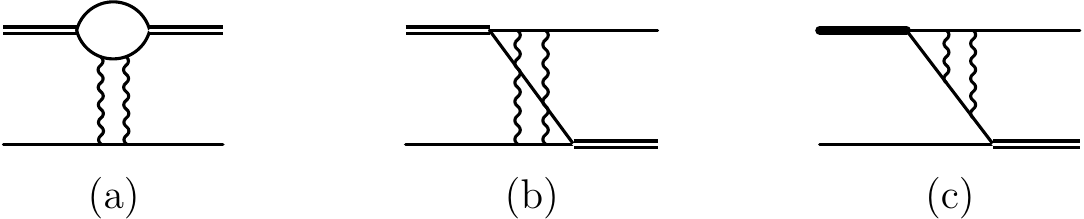}
\caption{Two-photon three-nucleon diagrams.  Solid lines represent nucleons 
whereas a deuteron (\OneSNot dibaryon) is drawn as a double (thick) line.  Wavy 
lines are Coulomb photons.}
\label{fig:TwoPhotonCoulomb}
\end{figure}

\begin{figure}[htbp]
\centering
\includegraphics[width=0.2\textwidth]{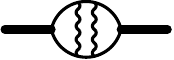}
\caption{Two-photon bubble.  Symbols are as in Fig.~\ref{fig:TwoPhotonCoulomb}.}
\label{fig:Corr-bub-double}
\end{figure}

As mentioned in the introduction, one motivation behind constructing the 
\OneSNot unitarity-limit scheme of Ref.~\cite{Konig:2015aka} was that it makes 
it very convenient to include perturbative Coulomb corrections in a way that 
ensures proper renormalization.  Specifically, the log-divergent diagram, where 
a Coulomb photon is exchanged inside a $pp$ bubble, is included at \NLO along 
with $\sigma_{t,pp}^{(1)}$, which absorbs the divergence when it is adjusted to 
give the physical (Coulomb-modified) $pp$ scattering length.  In this scheme, 
two-photon contributions (see Figs.~\ref{fig:TwoPhotonCoulomb} 
and~\ref{fig:Corr-bub-double}) enter at \NNLO.

%%%%%%%%%%%%%%%%%%%%%%%%%%%%%%%%%%%%%%%%%%%%%%%%%%%%%%%%%%%%%%%%%%%%%%%%%%%%%%
%%%%%%%%%%%%%%%%%%%%%%%%%%%%%%%%%%%%%%%%%%%%%%%%%%%%%%%%%%%%%%%%%%%%%%%%%%%%%%
%%%%%%%%%%%%%%%%%%%%%%%%%%%%%%%%%%%%%%%%%%%%%%%%%%%%%%%%%%%%%%%%%%%%%%%%%%%%%%

\begin{figure}[htbp]
\centering
\includegraphics[width=0.2\textwidth]{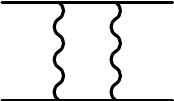}
\caption{General two-photon exchange diagram.  Symbols are as in 
Fig.~\ref{fig:TwoPhotonCoulomb}.}
\label{fig:TwoPhotonExchange}
\end{figure}

The basic ingredient for all these topologies is the two-photon loop diagram 
shown in Fig.~\ref{fig:TwoPhotonExchange}.  This could be calculated directly, 
but it is more convenient to simply extract it as the $\OO(\alpha^2)$ piece of 
the full off-shell Coulomb T-matrix.

%%%%%%%%%%%%%%%%%%%%%%%%%%%%%%%%%%%%%%%%%%%%%%%%%%%%%%%%%%%%%%%%%%%%%%%%%%%%%%
%%%%%%%%%%%%%%%%%%%%%%%%%%%%%%%%%%%%%%%%%%%%%%%%%%%%%%%%%%%%%%%%%%%%%%%%%%%%%%

For general kinematics with incoming (outgoing) momentum $\vp$ ($\vq$) and 
center-of-mass energy $E = k^2\MN$, this T-matrix can be written in the Hostler 
form~\cite{Chen:1972ab,Hostler:1964aa,Hostler:1964ab}
\begin{equation}
 T_C(k;\vecp,\vecq) = \frac{\!4\pi\alpha}{(\vecp-\vecq)^2}
 \left\{1-2\ii\eta
 \int_1^\infty\left(\frac{s+1}{s-1}\right)^{\!\!-\ii\eta}
 \frac{\dd s}{s^2-1-\epsilon} \right\}
 \mathtext{,}
 \epsilon = \frac{(p^2-k^2)(q^2-k^2)}{k^2(\vecp-\vecq)^2}
 \,,
\label{eq:TC-int}
\end{equation}
where
\begin{equation}
 \eta(k) = \frac{\alpha\MN}{2k}
\label{eq:eta-k}
\end{equation}
This form is most suitable for extracting the two-photon piece because it is 
straightforward to expand in $\alpha$.  Noting that $\eta=\OO(\alpha)$ gives
\begin{equation}
 T_C^{(2)}(k;\vecp,\vecq) = ({-}2\ii\eta)\frac{4\pi\alpha}{(\vecp-\vecq)^2}
 \int_1^\infty \frac{\dd s}{s^2-1-\epsilon}
 = {-}\ii \frac{4\pi\alpha^2\MN}{k(\vecp-\vecq)^2}
 \frac{\atanh\left(\sqrt{1+\epsilon}\right)}{\sqrt{1+\epsilon}} \,.
\label{eq:TC2}
\end{equation}
A numerical comparison with the bare expression obtained from 
Fig.~\ref{fig:TwoPhotonExchange}---which involves a brute-force 
three-dimensional momentum integration---gives excellent agreement with 
Eq.~\eqref{eq:TC2}.

The diagrams shown in Fig.~\ref{fig:TwoPhotonCoulomb} can now be expressed in 
terms of $T_C^{(2)}$, based on previous results including the full Coulomb 
T-matrix~\cite{Konig:2014ufa} (\cf~also 
Refs.~\cite{Kok:1979aa,Kok:1981aa,Ando:2010wq}).  Explicit expressions are 
given in Appendix~\ref{sec:CoulombKernels}.

Only the diagram with two photons exchange \emph{inside} a bubble, shown in 
Fig.~\ref{fig:Corr-bub-double}, requires some further work.  In principle, it 
can be obtained by integrating $T_C^{(2)}(k;\vecp,\vecq)$ over $\vp$ and $\vq$, 
but that would be unnecessarily tedious.  More conveniently, it can be 
extracted from the full Coulomb Green's 
function~\cite{Hostler:1963zz,Meixner:1933aa},
\begin{equation}
 G_C(E;\vecr'=\vZero,\vecr)
 = {-}2\mu\frac{\Gamma(1-\ii\eta)\,W_{\ii\eta;1/2}(-2\ii k r)}{4\pi r}
 \mathtext{,} \eta = \frac{\alpha\mu}{k}
 \mathtext{,} \mu = \frac{\MN}{2} \,.
\label{eq:GC-W}
\end{equation}
in the limit $\vecr\to0$.  When this is expanded in $\alpha$, the first to 
terms are divergent, but the $\OO(\alpha^2)$ and all higher orders are 
finite~\cite{Kong:1999sf,Kong:2000px}.  Expanding Eq.~\eqref{eq:GC-W} first in 
$r$ and subsequently in $\alpha$ gives\footnote{The logarithmic piece here 
actually comes together with a $\log r$ that has the same prefactor, and should 
thus be disregarded.}
\begin{equation}
 \frac{1}{2\mu} G_C(E;\vecr'=\vZero,\vecr) = \text{div.\ terms}
 - \frac{\ii k}{4\pi}
 + \frac{\alpha\mu\big[\log({-}2\ii k) + \EulerGamma - 1\big]}{2\pi}
 - \frac{\ii \mu^2 \pi \alpha^2}{12 k}
 + \OO(r,\alpha^3) \,,
\label{eq:GC-W-expansion}
\end{equation}
such that one can read off
\begin{equation}
 \delta J_0^{(2)}(k) = {-}\frac{\MN}{4\pi}
 \left[\frac{\ii\pi^2 \alpha^2\MN^2}{12k}\right] \,.
\label{eq:delta-J0-2}
\end{equation}
The label $\delta J_0^{(2)}(k)$ was chosen because alternatively this piece
can be extracted from the Coulomb bubble summing up two and more 
photon exchanges, which Ref.~\cite{Konig:2015aka} showed to be
\begin{equation}
 \delta J_0(k) = {-}\frac{\alpha\MN^2}{4\pi}
 \left[\psi(\ii\eta) + \frac{1}{2\ii\eta} + C_\Delta\right]
 + \frac{\MN}{4\pi} \ii k \,.
\label{eq:delta-J0}
\end{equation}
Expanding the digamma function as
\begin{equation}
 \psi(x) = {-}\frac{1}{x} - \EulerGamma + \frac{\pi^2x}{6} + \OO(x^2)
\end{equation}
gives again Eq.~\eqref{eq:delta-J0} and provides a consistency check.  In fact, 
all higher-order Coulomb contributions can be obtained in the same way as 
described here.

\subsection{Propagator corrections}
\label{sec:TwoBodyParamaters}

Next-to-leading order corrections to the dibaryon propagators are have been 
discussed in great detail in Ref.~\cite{Konig:2015aka}.  They are exactly the 
same here for the expansion that only takes the \OneSNot channel in the 
unitarity leading order.  For the ``full unitarity'' expansion introduced in 
Ref.~\cite{Konig:2016utl}, the \ThreeSOne channel is taken in the same limit at 
\LO, such that generically
\begin{subalign}[eq:Delta-dt-0-1]
 \ii\Delta_{d,t}^{(0)}(p_0,\vp)
 &= \frac{-\ii}{\sigma_{d,t}^{(0)} + y_{d,t}^2 I_0(p_0,\vp)} \\
 \ii\Delta_{t}^{(1)}(p_0,\vp)
 &= \ii\Delta_{d,t}^{(0)}(p_0,\vp)\times
 \bigg[{-}\ii\sigmadt^{(1)}
 - \ii \cdt^{(1)} \left(p_0-\frac{\vp^2}{4M_N}\right) \bigg]
 \times \ii\Delta_{d,t}^{(0)}(p_0,\vp) \,.
\end{subalign}
with
\begin{equation}
 \sigmadt^{(0)} = \frac{2\Lambda}{\pi}
 \mathtext{,}
 \sigmadt^{(1)} = {-}\frac{1}{a_{d,t}}
 \mathtext{,}
 \cdt^{(1)} = \frac{\MN r_{d,t}}{2} \,.
\label{eq:sigma-dt-unitarity}
\end{equation}
In particular, while in the standard and \OneSNot-unitarity \LO the \ThreeSOne 
propagator is matched to the effective range expansion around the deuteron 
pole, giving\footnote{The deuteron binding momentum is taken here as 
$\gamd=45.701~\MeV$~\cite{vanderLeun:1982aa} and the effective range is 
$\rd=1.765~\fm$~\cite{deSwart:1995ui}.}
\begin{equation}
 \sigmad^{(1)} = \frac{\rd\gamd^2}{2} \mathtext{,}
 \cd^{(1)} = \frac{\MN\rd}{2} \,,
\label{eq:sigmad1}
\end{equation}
the ordinary expansion around zero energy is used in the full-unitarity 
case to treat both channels fully equivalently.  Either way, at \NNLO there are 
quadratic insertions of $\sigmadt^{(1)}$ and $\cdt^{(1)}$,
\begin{equation}
 \ii\Delta_{t}^{(2)}(p_0,\vp)
 = \ii\Delta_{t}^{(0)}(p_0,\vp)\times
 \Bigg\{\bigg[{-}\ii\sigma_{t}^{(1)}
 - \ii \cdt^{(1)} \left(p_0-\frac{\vp^2}{4M_N}\right)\Bigg\}^{\!2} \,,
\label{eq:Delta-t-2}
\end{equation}
and of course this matches directly onto the expansion of the renormalized 
amplitude when the expressions from Eq.~\eqref{eq:sigma-dt-unitarity} are 
inserted.  Note that the generic dibaryon parameter $\sigma$ (in a given 
channel) is related the corresponding ``ordinary'' $C_0$ used in 
Sec.~\ref{sec:TwoBody-NNLO} via $\sigma = {-}{4\pi}/{(\MN C_0)}$.  This 
means that their formal expansions are different, and in general, for a fixed 
scattering length included at \NLO, one has $\sigma_{t(,pp)}^{(n)} = 0$ for $n 
\geq 2$.\footnote{The same is true for the \ThreeSOne channel expanded around 
zero momentum, but with the ERE around the deuteron pole, there are 
contributions to $\sigma_d$ from all orders.}  In the $nd$ sector, however, 
there is a correction $\sigma_{t,nn}^{(2)} = {-}1/\ann + 1/a_t$ which shifts 
the inverse scattering length away from its isospin-symmetric value.

\medskip
In the $pp$ channel, there are Coulomb contributions as well.  Most notably,
the bubble with a single photon exchanged inside---denoted as $\delta I_0(k)$---
is logarithmically divergent, and this divergence is absorbed by 
setting~\cite{Konig:2015aka}
\begin{equation}
 \sigma_{t,pp}^{(1)} = {-}\frac{1}{a_C}
 - \alpha\MN\left(\log\dfrac{2\Lambda}{\alpha\MN}
 - C_\zeta - C_\Delta\right)
\label{eq:renorm-sigma-t-pp-1}
\end{equation}
with known constants $C_\zeta$ and $C_\Delta$.\footnote{As discussed in detail 
in Ref.~\cite{Konig:2015aka}, the term $C_\zeta$ cancels exactly against the 
contribution from $\delta I_0(k)$, leaving only $C_\Delta\approx0.579$ in the 
final expressions.}  \NLO also includes isospin-symmetric range corrections 
given by $c_t^{(1)}$.  At \NNLO, there are quadratic insertions of 
$\sigma_{t,pp}^{(1)}$, $c_t^{(1)}$, and $\delta I_0(k)$, but in addition also 
the isospin-breaking range correction $c_{t,pp}^{(2)} = \MN(\rpp-r_t)/2$ and the 
genuine two-photon contribution $\delta J_0^{(2)}(k)$ discussed in the previous 
section.  Overall, this amounts to
\begin{spliteq}
 \ii\Delta_{t,pp}^{(2)}(p_0,\vp)
 &= \ii\Delta_{t,pp}^{(0)}(p_0,\vp)\times
 \Bigg\{\bigg[{-}\ii\sigma_{t,pp}^{(1)} 
 - \ii c_t^{(1)} \left(p_0-\frac{\vp^2}{4M_N}\right) \\
 &\hspace{9em}\null
 - \ii \yt^2\delta I_0\!\left(\ii\sqrt{\vp^2/4-\MN p_0-\ii\eps}\right)\bigg]
 \times \ii\Delta_{t,pp}^{(0)}(p_0,\vp)\Bigg\}^{\!2} \\
 &\null + \ii\Delta_{t,pp}^{(0)}(p_0,\vp)\times\bigg[
  {-}\ii c_{t,pp}^{(2)} \left(p_0-\frac{\vp^2}{4M_N}\right) \\
 &\hspace{9em}\null
  - \ii\yt^2\delta J_0^{(2)}\!\left(\ii\sqrt{\vp^2/4-\MN p_0-\ii\eps}\right)
 \bigg]\times \ii\Delta_{t,pp}^{(0)}(p_0,\vp) \,,
\label{eq:Delta-pp-N2LO-1}
\end{spliteq}
and inserting the various definitions this simplifies to
\begin{spliteq}
 \ii\Delta_{t,pp}^{(2)}(p_0,\vp) &= {-}\ii\dfrac{
 \Bigg\{\dfrac{1}{\app} - \alpha\MN\left[C_\Delta
 + \log\!\left(\dfrac{\alpha\MN}{2\tilde{k}(p_0,\vp)}\right)
 \right] + \dfrac{r_t}{2}\,\tilde{k}(p_0,\vp)^2 \Bigg\}^{\!2}
 - \dfrac{\pi^2\alpha^2\MN^2}{12}}
 {\tilde{k}(p_0,\vp)^3} \\
 &\null - \ii \frac{\rpp - r_t}{2} \,,
\label{eq:Delta-pp-N2LO-2}
\end{spliteq}
where $\tilde{k}(p_0,\vp) \equiv \sqrt{{\vp^2}/{4}-\MN p_0-\ii\eps}$.

\section{Three-body observables up to \NNLO}
\label{sec:ThreeBody}

\subsection{Integral equations}
\label{sec:ThreeBodyEquations}

Adapting the notation of Refs.~\cite{Konig:2014ufa,Konig:2015aka}, the LO 
amplitude is a three-vector in channel space (with the last two components 
corresponding to $np$ and $pp/nn$ singlet-dibaryon legs):
\begin{equation}
 \vTS \equiv \left(\TSda,\TSdb,\TSdc\right)^T \,.
\end{equation}
It is determined by the integral equation
\begin{equation}
 \vTS^{(0)}
 = {-}\hat{K}^{(0)}\vec{e}_1
 + (\hat{K}^{(0)}\hat{D}^{(0)}) \otimes \vTS^{(0)} \,,
\label{eq:Nd-IntEq-0}
\end{equation}
where $\vec{e}_1=(1,0,0)^T$ and $\otimes$ represents an integral over the 
intermediate momentum,
\begin{equation}
 A \otimes B \equiv \frac1{2\pi^2}
 \int_0^\Lambda\dd q\,q^2\,A(\ldots,q)B(q,\ldots) \,,
\label{eq:SH-int}
\end{equation}
involving a momentum cutoff $\Lambda$.  $\hat{D}^{(0)})$ is a diagonal matrix of 
dibaryon propagators, \viz
\begin{equation}
 \hat{D}^{(n)} = \diag\left(D_d^{(n)},D_t^{(n)},D_{t(,pp)}^{(n)}\right) \,,
\label{eq:D-n}
\end{equation}
where generically $D_x^{(n)}(E;q) \equiv 
\Delta_x^{(n)}\big(E-q^2/(2\MN);q\big)$, and the 
kernel matrix is
\begin{equation}
 \hat{K}^{(0)} = 2\pi \begin{pmatrix}
 {-}\left(\KS+\frac{2H_{0,0}(\Lambda)}{\Lambda^2}\right) &
 \left(3\KS+\frac{2H_{0,0}(\Lambda)}{\Lambda^2}\right) &
 \left(3\KS+\frac{2H_{0,0}(\Lambda)}{\Lambda^2}\right)\\[\skrowspace]
 \left(\KS+\frac{2H_{0,0}(\Lambda)}{3\Lambda^2}\right) &
 \left(\KS-\frac{2H_{0,0}(\Lambda)}{3\Lambda^2}\right) &
 {-}\left(\KS+\frac{2H_{0,0}(\Lambda)}{3\Lambda^2}\right)\\[\skrowspace]
 \left(2\KS+\frac{4H_{0,0}(\Lambda)}{3\Lambda^2}\right) &
 {-}\left(2\KS+\frac{4H_{0,0}(\Lambda)}{3\Lambda^2}\right) &
 {-}\frac{4H_{0,0}(\Lambda)}{3\Lambda^2}
 \end{pmatrix} \,.
\label{eq:K-0}
\end{equation}
It describes the isospin-symmetric system without Coulomb contributions and 
thus only contains the one-nucleon exchange part $\KS$ and the \LO 
three-nucleon force $H_{0,0}$.  For more detailed definitions, see 
Ref.~\cite{Konig:2014ufa}.

As indicated by the subscript ``$\sss$,'' the \LO equation only contains 
contributions from the strong interaction.  The additional inclusion of Coulomb 
effects perturbatively builds up the ``full'' amplitude.  Up to \NNLO, the
corresponding integral equations are
\begin{equation}
 \vTF^{(1)}
 = {-}\hat{K}^{(1)}\vec{e}_1
 + \left[\hat{K}^{(1)}\hat{D}^{(0)} + \hat{K}^{(0)}\hat{D}^{(1)} \right]
  \otimes \vTF^{(0)}
 + (\hat{K}^{(0)}\hat{D}^{(0)}) \otimes \vTF^{(1)}
\label{eq:Nd-IntEq-1}
\end{equation}

and
\begin{multline}
 \vTF^{(2)}
 = {-}\hat{K}^{(2)}\vec{e}_1
 + \left[\hat{K}^{(2)}\hat{D}^{(0)}
 + \hat{K}^{(1)}\hat{D}^{(1)} + \hat{K}^{(0)}\hat{D}^{(2)}
 \right] \otimes \vTF^{(0)} \\
 + \left[\hat{K}^{(1)}\hat{D}^{(0)} + \hat{K}^{(0)}\hat{D}^{(1)} \right]
  \otimes \vTF^{(1)}
 + (\hat{K}^{(0)}\hat{D}^{(0)}) \otimes \vTF^{(2)} \,,
\label{eq:Nd-IntEq-2}
\end{multline}
with $\vTF^{(0)} = \vec{\TS}^{(0)}$.  Essentially, the structure is the 
same as in the two-body example discussed in Sec.~\ref{sec:TwoBody-NNLO}, only 
that here there are two types of corrections:
\begin{enumerate}
\item Contributions to the dibaryon propagators (from finite scattering 
lengths and effective ranges as well as two-body Coulomb effects in the $pp$ 
subsystem), encoded in the general expression given in Eq.~\eqref{eq:D-n}.
\item  The associated three-body force corrections as well as three-body Coulomb 
diagrams contributing to the higher-order interaction-kernel matrices.
\end{enumerate}
Specifically, these matrices are
\begin{equation}
 \hat{K}^{(1)} = 2\pi \begin{pmatrix}
 {-}\left(\Kbub+\Kbox+\frac{2H_{0,1}(\Lambda)}{\Lambda^2}\right) &
 \left(3\Kbox+\frac{2H_{0,1}(\Lambda)}{\Lambda^2}\right) &
 \left(3\KtriOut+\frac{2H_{0,1}(\Lambda)}{\Lambda^2}\right)\\[\skrowspace]
 \left(\Kbox+\frac{2H_{0,1}(\Lambda)}{3\Lambda^2}\right) &
 {-}\left(\Kbub-\Kbox+\frac{2H_{0,1}(\Lambda)}{3\Lambda^2}\right) &
 {-}\left(\KtriOut+\frac{2H_{0,1}(\Lambda)}{3\Lambda^2}\right)\\[\skrowspace]
 \left(2\KtriIn+\frac{4H_{0,1}(\Lambda)}{3\Lambda^2}\right) &
 {-}\left(2\KtriIn+\frac{4H_{0,1}(\Lambda)}{3\Lambda^2}\right) &
 {-}\frac{4H_{0,1}(\Lambda)}{3\Lambda^2}
 \end{pmatrix}
\label{eq:K-1}
\end{equation}
and
\begin{multline}
 \hat{K}^{(2)} = 2\pi \begin{pmatrix}
 {-}\left(\KbubTwo+\KboxTwo+2H^{(2)}\right) &
 \left(3\Kbox+2H^{(2)}\right) &
 \left(3\KtriOutTwo+2H^{(2)}\right)\\[\skrowspace]
 \left(\KboxTwo+\frac23H^{(2)}\right) &
 {-}\left(\KbubTwo-\KboxTwo+\frac23H^{(2)}\right) &
 {-}\left(\KtriOutTwo+\frac23H^{(2)}\right)\\[\skrowspace]
 \left(2\KtriInTwo+\frac43H^{(2)}\right) &
 {-}\left(2\KtriInTwo+\frac43H^{(2)}\right) &
 {-}\frac43H^{(2)}
 \end{pmatrix} \\
 \null + 2\pi \, \diag\big(\Krd,\Krt,0\big) \,,
\label{eq:K-2}
\end{multline}
where the collection of \NNLO three-nucleon forces is
\begin{equation}
 H^{(2)}(\Lambda) = \frac{H_{0,2}(\Lambda)}{\Lambda^2}
 + \frac{H_{2,2}(\Lambda)}{\Lambda^4}(\MN E + \gamd^2)
 + \frac{H_{0,2}^{(\alpha)}(\Lambda)}{\Lambda^2} \,.
\end{equation}

The functions $\KbubTwo$, $\KboxTwo$, and $\KtriOutTwo$ correspond, 
respectively, to the three diagrams shown in Fig.~\ref{fig:TwoPhotonCoulomb}.  
$\KtriInTwo$ comes from the reversed version of 
Fig.~\ref{fig:TwoPhotonCoulomb}(c), where the photon exchanges are on the left 
side of the diagram.  Detailed expressions for all these functions are given in 
the Appendix~\ref{sec:CoulombKernels}.  The analogous one-photon functions in 
Eq.~\eqref{eq:K-1} are defined (for example) in Ref.~\cite{Konig:2015aka} and 
are thus not repeated here.  Finally, the contributions $\Krd$ and $\Krt$, 
corresponding to a direct coupling of photons to the dibaryons,
\begin{equation}
 K_{r}(E;k,p) = {-}\frac{\alpha\MN r}{2kp}
 \,Q_0\!\left(-\frac{k^2+p^2+\lambda^2}{2kp} \right)
 \mathtext{,}
 Q_0(x) = \frac{1}{2}\log\left(\frac{x+1}{x-1}\right)
 \mathtext{,}
 r = \rho_d, r_t \,,
\label{eq:K_r}
\end{equation}
enter at \NNLO in the power counting ($\lambda$ here is a small photon mass 
introduced as infrared regulator and discussed further in the next section).

Note that the new three-body force $H_{0,2}^{(\alpha)}(\Lambda)$ is needed 
due to terms mixing range corrections and $O(\alpha)$ Coulomb contributions.
While from the \NNLO calculation based on the T-matrix method it is not 
directly obvious, one can identify contributions to the energy shift of the form
\begin{equation}
 \vcenteredhbox{%
  \raisebox{0.2em}{\includegraphics[width=5em]{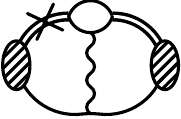}}
 }
 \sim \; \left(\frac1q\right)^{\!2}_{\!\!\Bgen}
 \times \Big(q^3\Big)^{3}_{\!\text{loops}}
 \times \left(\frac1{q^2}\right)^{\!2}_{\!\!N}
 \times \left(\frac1q\right)^{\!3}_{\!\!D}
 \times \left(\frac1{q^2}\right)_{\!\text{photon}}
 \times \Big(q^2\Big)_{\!\text{range corr.}}
 \;\sim\; q^0 \,,
\end{equation}
where all loop momenta have been generically denoted by $q$ and the notation of 
Ref.~\cite{Konig:2014ufa} has been adapted to count ultraviolet behavior 
diagram.  This, as well as similar diagrams involving other topologies and/or 
a direct dibaryon-photon coupling, are logarithmically divergent, such that the 
three-body forces determined from the $nd$ system alone are no longer sufficient 
to ensure renormalization.

\subsection{Extraction of binding energies}

Binding energies can be obtained by plugging the solutions of the integral 
equations into the expressions given in Sec.~\ref{sec:SecondOrderEnergies}.
At LO, this merely amounts to varying the energy until the homogeneous version 
of Eq.~\eqref{eq:Nd-IntEq-0} has a solution.  When the \LO three-nucleon force 
is fitted to the physical triton binding energy, the energy is fixed at that 
value and $H_{0,0}(\Lambda)$ is varied (for a given value of $\Lambda$).

The perturbative corrections to the binding energy are defined in terms of 
limits where the energy approaches the \LO pole.  Numerically, this is evaluated 
by varying the energy around the pole and interpolating with a low-order 
polynomial to extract the residues (a linear interpolation is found to be 
sufficient in most cases if a 1\% interval around the pole is considered).
The reliability of this procedure has been checked by comparing \NLO results 
with those obtained as matrix elements between leading-order vertex functions 
(as described, for example, in Ref.~\cite{Konig:2015aka}), finding excellent 
agreement.

\subsection{Subtracted phase shifts}

For $pd$ scattering it is necessary to calculate Coulomb-subtracted phase 
shifts $\tilde{\delta}(k) \equiv \delta_\fff(k) - \delta_\ccc(k)$, where the 
first term---corresponding to the combination of strong and Coulomb 
contributions---is extracted perturbatively from the full amplitude 
(specifically, from the the $pd \to pd$ component $\TF\equiv\TFda$).  Up to 
\NNLO, the expressions are
\begin{subalign}
 \delta_{\fff}^{(0)}(k)
 &= \frac{1}{2\ii} \log\!\left(1+\tfrac{\ii k\mu}{\pi}
 T_\fff^{(0)}(k)\right) \,, \\
 \delta_{\fff}^{(1)}(k) &= \frac{k\mu}{2\pi}
 \dfrac{T_\fff^{(1)}(k)}{1+\tfrac{\ii k\mu}{\pi} T_\fff^{(0)}(k)} \,, \\
 \delta_{\fff}^{(2)}(k) &= \frac{k\mu}{2\pi}
 \dfrac{T_\fff^{(2)}(k)}{1+\tfrac{\ii k\mu}{\pi} T_\fff^{(0)}(k)}
 - \ii \delta_{\fff}^{(1)}(k)^2 \,,
\end{subalign}
where the T-matrices, in turn, combine the perturbative expansion of the 
deuteron wavefunction renormalization,
\begin{equation}
 Z_0^{-1} = \ii\frac{\partial}{\partial p_0}
 \left.\frac{1}{\Delta_d(p_0,\vp)}\right|_{p_0 =-\frac{\gamd^2}{\MN},\,\vp=0}
 \implies Z_0 = Z_0^{(0)} + Z_0^{(1)} + \cdots \,,
\label{eq:Z0}
\end{equation}
with that of the bare amplitude, \viz
\begin{equation}
 T_\fff^{(0)}(k) = Z_0^{(0)}\TF^{(0)}(E_k;k,k)
 \mathtext{,}
 T_\fff^{(1)}(k) = Z_0^{(0)}\TF^{(1)}(E_k;k,k)+Z_0^{(1)}\TF^{(0)}(E_k;k,k) 
\,,
\end{equation}
and following the same pattern for $\TF^{(2)}(k)$.  As before, the reduced mass 
is $\mu=2\MN/3$.

\begin{figure}[tb]
\centering
\includegraphics[width=0.75\textwidth,clip]{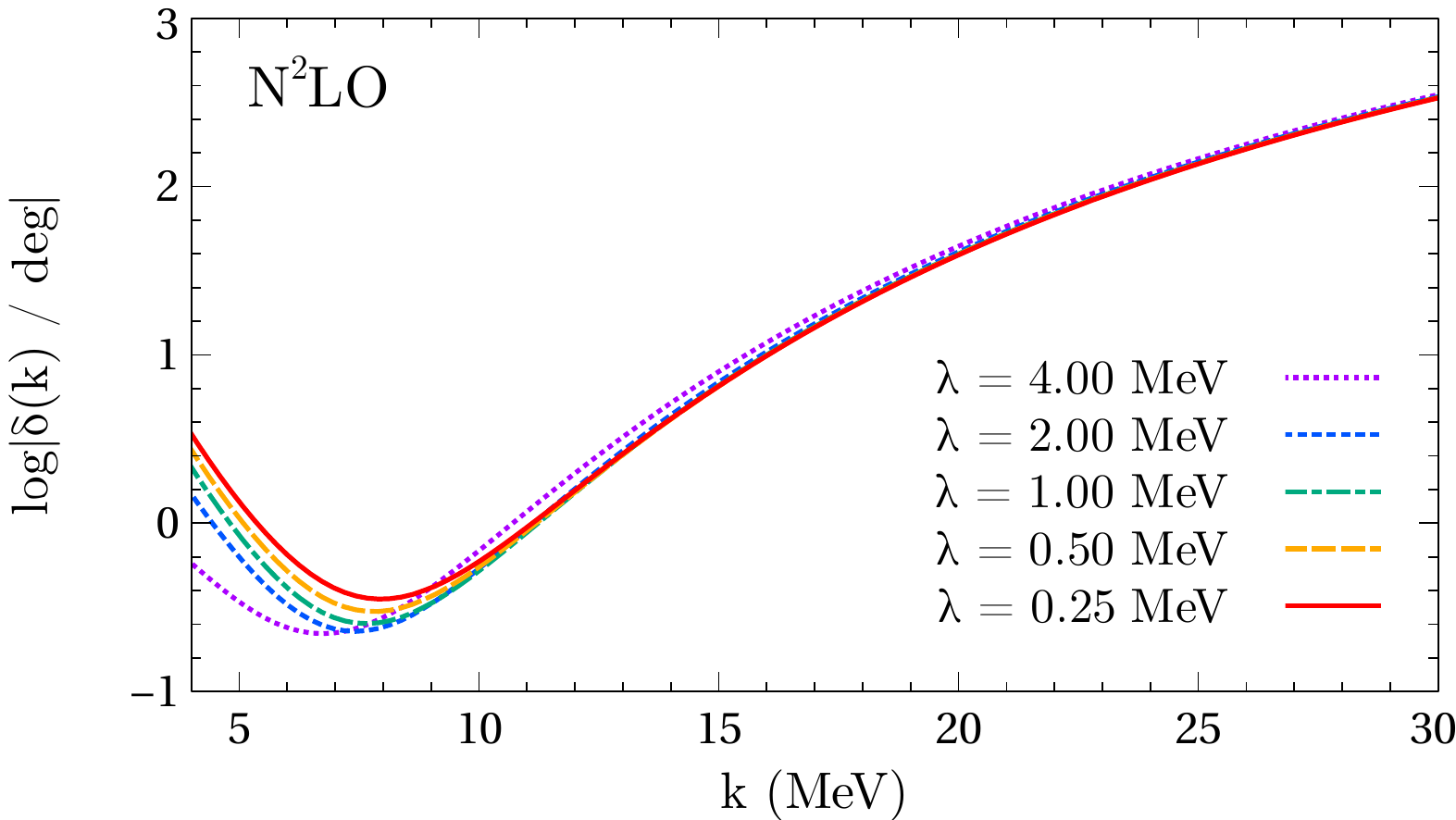}
\caption{%
Proton-deuteron S-wave phase shifts calculated at \NNLO for different values of 
the regulating photon mass $\lambda$.  The logarithm is taken to enhance 
the visibility of small differences, and overall convergence as $\lambda$ is 
lowered towards zero is apparent.  All curves were calculated at a cutoff 
$\Lambda=4.8~\GeV$}
\label{fig:Phase-pd-U-2-lambda}
\end{figure}

Most Coulomb contributions are formally divergent in the on-shell limit.  This 
infrared singularity corresponds to the long-range nature of the Coulomb force 
and is conveniently regulated by introducing a small photon mass $\lambda$ 
where necessary, as it was done for example in Eq.~\eqref{eq:K_r}.  While in 
the low-energy limit, where Coulomb effects are highly nonperturbative, this 
``screening'' is very important~\cite{Konig:2013cia}, in the perturbative 
regimes considered here (\ThreeHe and $pd$ scattering at intermediate 
center-of-mass momenta), there is almost no noticeable $\lambda$ 
dependence in the subtracted phase shifts for photon masses below about 
$1~\MeV$ (\cf~results shown in Fig.~\ref{fig:Phase-pd-U-2-lambda}). In practice,
a fixed small value can be used instead of numerically taking the limit
$\lambda\to0$.

According to Ref.~\cite{Konig:2013cia}, the pure Coulomb part should be taken 
in a simple two-body picture in order to compare to experimental phase-shift 
analyses and potential-model calculations.   Including the photon mass, this 
corresponds to a Yukawa interaction between proton and deuteron.  At \LO, there 
are no Coulomb effects by construction, so $T_\ccc^{(0)}(k)=0$ and 
$\delta_{\ccc}^{(0)}(k) = 0$.  At \NLO and \NNLO, exact expressions can be 
obtained from the analogous case of pion-exchange contributions in the \OneSNot 
nucleon-nucleon channel.  Adapting results obtained 
in Ref.~\cite{Fleming:1999ee} for pion-exchange diagrams to the 
formally analogous $pd$-Coulomb case gives
\begin{equation}
 \delta_\ccc^{(1)}(k) = \frac{k\mu}{2\pi} T_\ccc^{(1)}(k)
 = {-}\frac{\alpha\mu}{2k}\log\left(1+\frac{4k^2}{\lambda^2}\right)
\end{equation}
and $\delta_\ccc^{(2)}(k) = k\mu\,T_\ccc^{(2)}(k)/(2\pi) - 
\ii\delta_\ccc^{(1)}(k)^2$
with
\begin{equation}
 T_\ccc^{(2)}(k) = \frac{2\mu\alpha^2\pi}{k^3}\left[
 \frac{\ii}{4}\log\left(1+\frac{4k^2}{\lambda^2}\right)^{\!2}
 + \Ip\,\LiTwo\!\left(\frac{2k^2 - \ii k\lambda}{\lambda^2+4k^2}\right)
 + \Ip\,\LiTwo\!\left(\frac{{-}2k^2 + \ii k\lambda}{\lambda^2}\right)
 \right] \,.
\end{equation}

\section{Results}
\label{sec:Results}

\subsection{The $\boldsymbol{n}\boldsymbol{d}$ system}

As a first step, the \OneSNot-unitarity expansion can be tested at \NNLO in the 
$nd$ system.  The corresponding amplitudes are readily obtained by omitting all 
Coulomb contributions in the integral equations discussed in 
Sec.~\ref{sec:ThreeBodyEquations} and changing the $pp$ to an $nn$ propagator.  
The $nd$ doublet scattering length can then be calculated as
\begin{equation}
 \andDoublet = {-}\frac{\MN}{3\pi}\lim\nolimits_{k\to0} Z_0\TSda(E_k;k,k) \,,
\label{eq:and-2}
\end{equation}
with appropriate perturbative expressions at each order.

While Ref.~\cite{Konig:2015aka} found that already at \NLO in the $1/|a_t|$ 
expansion, results for both the Phillips line (\ie, the correlation between 
$\andDoublet$ and the \ThreeH binding energy~\cite{Phillips:1968zze}) as well 
as $nd$ phase shifts fall almost on top of those in the standard pionless 
counting (including $1/a_t$ nonperturbatively at \LO), the agreement is even 
better at \NNLO (relative differences compared to standard \LO are an 
order of magnitude smaller at \NNLO than at \NLO, of the order $10^{-5}$ for 
the Phillips line and $10^{-4}$ for the $nd$ phase shift), thus establishing 
more firmly that the expansion converges rapidly.

\subsection{${}^{\boldsymbol{3}}\mathbf{He}$ binding energy}
\label{sec:ThreeHeResults}

Studying the \ThreeHe bound state is more interesting.  Setting all range 
corrections to zero, it is possible to compare the expansions suggested, 
respectively, in Refs.~\cite{Konig:2015aka} and~\cite{Konig:2016utl} at \NNLO.  
In the first case, only the \OneSNot channel is expanded around the unitarity 
limit, whereas the latter does the same for the \ThreeSOne channel.  Since 
$a_d\sim5.4~\fm < |a_t|\sim23.7~\fm$, corrections are expected to be larger in 
the full-unitarity case.

%%%%%%%%%%%%%%%%%%%%%%%%%%%%%%%%%%%%%%%%%%%%%%%%%%%%%%%%%%%%%%%%%%%%%%%%%%%%%%
\begin{figure}[tb]
\centering
\includegraphics[width=0.75\textwidth,clip]{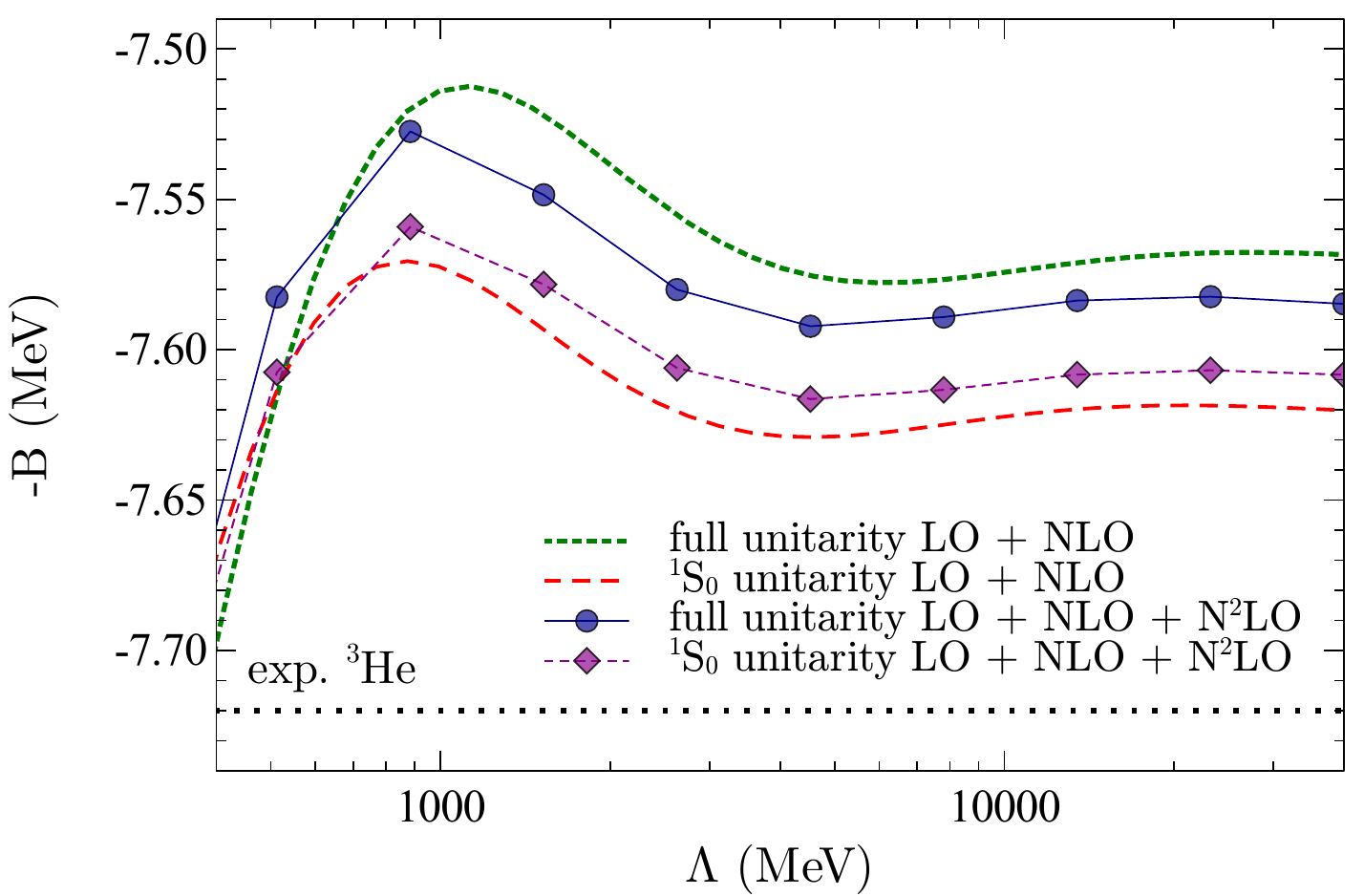}
\caption{%
Three-nucleon binding energies up to second-order in the scattering-length 
expansion (\ie, without effective-range corrections).  The (red) dashed curve 
and diamonds show results for the expansion of Ref.~\cite{Konig:2015aka}, which 
takes only the \OneSNot channel in the unitarity limit at \LO.  The (green) 
dotted line and circles use the ``full-unitarity'' scheme of 
Ref.~\cite{Konig:2016utl}, which takes the \ThreeSOne channel in the 
\LO-unitarity limit as well.  For each calculation, the isospin-symmetric 
three-nucleon forces have been fitted to keep the triton binding energy fixed 
at its physical value.  The horizontal dotted line indicates the experimental 
value for the \ThreeHe binding energy.}
\label{fig:En-3He-UU-2}
\end{figure}
%%%%%%%%%%%%%%%%%%%%%%%%%%%%%%%%%%%%%%%%%%%%%%%%%%%%%%%%%%%%%%%%%%%%%%%%%%%%%%

Results are shown in Fig.~\ref{fig:En-3He-UU-2}.  Note that convergence towards 
the experimental value is not to be expected in this expansion that (besides 
Coulomb effects) only perturbatively includes the finite scattering length(s).  
What is important, however, is that the results for full and \OneSNot-only 
unitarity move closer together at \NNLO, which indicates that both 
expansions---and in particular the perturbative treatment of $1/a_d$---are 
indeed convergent.

While for \OneSNot unitarity there is a little less binding at \NNLO compared 
to \NLO, the effect is opposite for the full-unitarity \LO case.  This can 
generally be understood as a cancellations between different contributions at 
this order, which play out slightly differently in the two cases.  Starting at 
\NLO in the full-unitarity scheme, $1/a_d$ corrections add attraction (in the 
unitarity limit, there is less attraction in the \ThreeSOne channel compared to 
finite $a_d>0$), requiring a repulsive contribution to the three-nucleon force 
to keep the triton in place.  The opposite effect happens in the \OneSNot 
channel (where the physical scattering length is negative), but it is smaller 
compared to the triplet contribution since $1/|a_t| < 1/a_d$.  Coulomb effects 
and $1/a_C<0$ provide pure repulsion at \NLO, and this is the only net effect 
at that order because all isospin-symmetric contributions cancel when the 
\ThreeH--\ThreeHe binding-energy difference is calculated in first-order 
perturbation theory.  At \NNLO there are nonlinear (quadratic) mixtures of the 
various \NLO insertions, as well as genuine \NNLO terms (two-photon diagrams 
and the \ann correction) that enter linearly at this order, so that there is no 
clear \apriori intuition what the net effect should be.  In any case, the 
overall corrections are small (note the zoomed scale in 
Fig.~\ref{fig:En-3He-UU-2}), as one would expect at \NNLO in this expansion.

\medskip
Range corrections are generally an important effect, but they cancel at \NLO 
for the same reason as discussed above (adjustment of the three-nucleon force). 
At \NNLO, their inclusion gives rise to divergences, stemming both directly 
from isospin breaking in the effective ranges, $r_C \neq r_t$ as well as from 
the interference of (isospin-symmetric) range corrections and Coulomb 
contributions that entered together at \NLO.  An isospin breaking three-nucleon 
force is thus required at this order for renormalization (just like the \NLO 
term identified in Ref.~\cite{Vanasse:2014kxa} for standard pionless counting 
with Coulomb effects included nonperturbatively), and it is no longer possible 
to determine the \ThreeHe binding energy without input from a $pd$ doublet 
observable.  This input can be chosen to be the \ThreeHe energy itself, 
allowing a prediction of $pd$ scattering.

\subsection{$\boldsymbol{p}\boldsymbol{d}$ scattering}
\label{sec:pdResults}

\begin{figure}[tb]
\centering
\includegraphics[width=0.75\textwidth,clip]{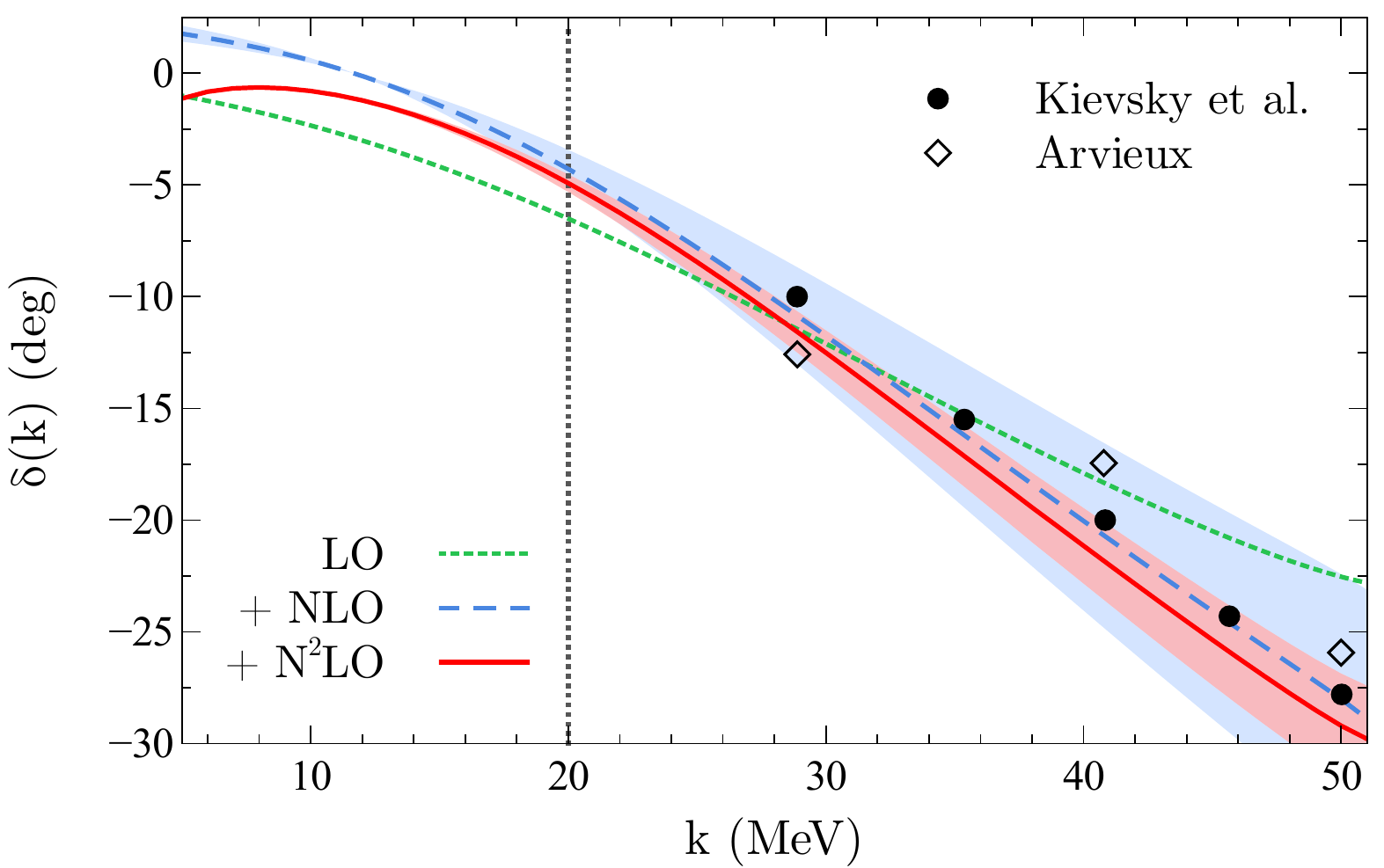}
\caption{%
Proton-deuteron S-wave phase shifts calculated up to \NNLO, with the \LO 
calculation identical to the $nd$ result and Coulomb and other isospin breaking 
effects added perturbatively at subsequent orders.  The isospin-breaking 
three-nucleon force has been fitted at to reproduce the experimental value for 
the \ThreeHe binding energy at \NNLO.  All curves were calculated at a cutoff 
$\Lambda=4.8~\GeV$ and a fixed photon mass $\lambda=0.25~\MeV$ has been used as 
infrared regulator.  As demonstrated in Fig.~\ref{fig:Phase-pd-U-2-lambda}, 
which shows the dependence of the \NNLO result, this is sufficiently close to the 
limit $\lambda\to0$ for practical purposes.  The filled circles were calculated 
in Ref.~\cite{Kievsky:1996ca} using the AV18+UIX potential model.  Experimental 
data (open diamonds) is shown from Ref.~\cite{Arvieux:1974fma}.}
\label{fig:Phase-pd-U}
\end{figure}

Scattering phase shifts for the $pd$ doublet channel are shown in 
Fig.~\ref{fig:Phase-pd-U}.  For this calculation, the three-nucleon 
force fitting strategy has been adjusted to match what was used in 
Ref.~\cite{Vanasse:2013sda}, \ie, instead of determining $H_{0,n}(\Lambda)$ by 
demanding that the triton remains at its physical energy, the $nd$ doublet 
scattering length---calculated according to Eq.~\eqref{eq:and-2}---is used as 
input instead.  The triton is then moved into place at \NNLO by adjusting 
$H_{2,2}(\Lambda)$.\footnote{At \NLO, it is also possible to calculate \ThreeHe 
with $H_{0,n}(\Lambda)$ fitted to the $nd$ scattering length.  While this 
would move the absolute position of the predicted binding energy, the 
difference to the triton---which is what is really calculated---remains 
unchanged.}  On top of this, $H_{0,2}^{(\alpha)}(\Lambda)$ is finally adjusted 
to reproduce the \ThreeHe energy.  Following this procedure has the advantage 
that all terms can be determined independently in subsequent one-parameter fits. 
 Note that this calculation is carried out for the \OneSNot-unitarity scheme 
only in order to have a fixed deuteron bound state.

%%%%%%%%%%%%%%%%%%%%%%%%%%%%%%%%%%%%%%%%%%%%%%%%%%%%%%%%%%%%%%%%%%%%%%%%%%%%%%
%%%%%%%%%%%%%%%%%%%%%%%%%%%%%%%%%%%%%%%%%%%%%%%%%%%%%%%%%%%%%%%%%%%%%%%%%%%%%%
%%%%%%%%%%%%%%%%%%%%%%%%%%%%%%%%%%%%%%%%%%%%%%%%%%%%%%%%%%%%%%%%%%%%%%%%%%%%%%

The \LO result (green dotted curve in Fig.~\ref{fig:Phase-pd-U}) coincides 
with that for $nd$ scattering.  Since all Coulomb effects are included 
perturbatively here, the $pd$ calculation is not expected to work at very low 
center-of-mass momenta $k$, but already at $k=20~\MeV$ the Coulomb parameter 
$\alpha\MN/k$ is about $1/3$ (in fact, for the $pd$ system one should rather 
count $\alpha2\MN/(3k)$ to account for the proper reduced mass, and this is 
$\simeq1/4$ at $k=20~\MeV$).  Consequently, the perturbative calculation should 
work well for $k\gsim20~\MeV$, and higher-order Coulomb effects are certainly 
not the primary source of uncertainty at \NNLO.  Following 
Ref.~\cite{Konig:2015aka}, the $Q/\LamNoPi$ and $\aleph_0/Q$ expansions are 
paired by taking $Q\sim\sqrt{\mathstrut\aleph_0\LamNoPi}$.  Conservatively 
choosing $\aleph_0 \sim 1/\app$ (\ie, assuming it is dominated by the largest 
scale) and $\LamNoPi \sim \Mpi$ gives $Q\sim60~\MeV$, slightly above the 
standard pionless counting which generally assumes $Q\sim\gamd\sim50~\MeV$.  
With this, the uncertainties at \NLO and \NNLO are estimated, respectively, as 
$Q((\aleph_0/Q)^2) \simeq 20\%$ and $Q((\aleph_0/Q)^3) \simeq 8\%$.  They are 
indicated as shaded bands in Fig.~\ref{fig:Phase-pd-U}.  Within these bands, 
potential-model results from Ref.~\cite{Kievsky:1996ca} (using the AV18+UIX 
interaction) are accommodated.  Results from an experimental phase-shift 
analysis~\cite{Arvieux:1974fma} are shown as well.  While overall the agreement 
looks reasonable, a more rigorous statement cannot be made in the absence of 
error bars for the data points.

\section{Summary and outlook}
\label{sec:Conclusion}

This work demonstrates that the pionless counting schemes developed in 
Refs.~\cite{Konig:2015aka} and~\cite{Konig:2016utl}, which take, respectively, 
the \OneSNot or both $NN$ S-wave channels in the unitarity limit at leading 
order, work well up to second-order perturbation theory.  The inclusion of 
effective-range corrections requires an isospin-breaking three-nucleon force in 
the doublet channel at \NNLO, but fixing this to a single datum (like the 
\ThreeHe binding energy) allows predictions to be made at other energies.  In 
particular, there exists a regime where $pd$ scattering is well described using 
only perturbative Coulomb effects (with two-photon contributions included at 
\NNLO).  Within the assigned \apriori uncertainty based on the EFT expansion, 
results agree with potential-model calculations of $pd$ scattering.  This 
establishes that \NNLO calculations of the nuclear three-body system including 
Coulomb effects are feasible and well-behaved, which is an important proof of 
principle on the way to similar calculations to be carried out in chiral EFT.

It is clear that some nonperturbative treatment of Coulomb effects is 
needed as $k\to0$ in the three-nucleon system, \ie, for a calculation of the 
$pd$ scattering length.  At the very least, the diagram with a single photon 
attached to the bubble should be iterated at \LO to reflect its infrared 
enhancement at small energies.  For the quartet channel (total spin 
$\nicefrac32$), exactly this was done in Ref.~\cite{Konig:2013cia}, with the box 
diagram and the dibaryon-photon range coupling added perturbatively (note 
that treating Coulomb diagrams differently based on which momentum 
scale dominates their behavior is generally what has been suggested in 
Ref.~\cite{Rupak:2001ci}).  Extending this procedure to the doublet channel, 
using the \OneSNot-unitarity expansion to handle the Coulomb divergence in the 
$pp$ channel, is left for future work, as is an analysis of Coulomb effects in 
$pd$ scattering above the deuteron breakup threshold ($k\simeq53~\MeV$), where 
the energy in the two-body subsystem goes through zero.  Moreover, it will be 
interesting for future work to combine the unitarity-\LO expansions employed 
here with the method developed in Ref.~\cite{Vanasse:2015fph} in order to 
calculate the \ThreeHe charge radius.

Typical for pionless EFT, the calculation carried out here allows for its 
various low-energy constants to be determined independently.  This is done
analytically in the two-nucleon sector and using subsequent one-parameter fits 
for the $3N$ forces, which is conveniently implemented numerically.  However, by 
not taking into account the global uncertainty at a given order, one effectively 
assumes some higher-order terms not to contribute.  In Ref.~\cite{Konig:2015aka} 
it was already noted that by fixing the $pp$ scattering length exactly to its 
experimental value at \NLO one overestimates the role of Coulomb effects over 
short-range isospin-breaking contributions.  Something similar happens by 
demanding that the \ThreeHe binding energy takes its exact physical value at 
\NNLO via the adjustment of the isospin-breaking three-nucleon force.  While of 
course such effects are reflected in the assigned uncertainty estimates, it 
would be interesting to carry out a more sophisticated fitting procedure that 
propagates the uncertainties and correlations through different orders and 
derives probability distributions for the low-energy constants.  Such methods 
are currently being developed, with the primary goal of applying them to chiral 
EFT (see for example Ref.~\cite{Wesolowski:2015fqa} and earlier references 
therein).  Pionless EFT provides a convenient test case for such a framework 
because results can be compared directly to explicitly determined values for the 
LECs.  Work along these lines is in progress.

\begin{acknowledgments}
I would like to thank R.~Furnstahl and U.~van Kolck for useful discussions and 
valuable comments on the manuscript.  Moreover, I am grateful to the 
participants of the GSI-funded EMMI RRTF workshop \emph{ER15-02: Systematic 
Treatment of the Coulomb Interaction in Few-Body Systems} for a very stimulating 
meeting.  This work is supported in part by the NSF under award PHY--1306250, 
by the DOE-funded NUCLEI SciDAC Collaboration under award DE-SC0008533, as well 
as by the ERC Grant No.\ 307986 STRONGINT.
\end{acknowledgments}

\appendix

\section{Two-photon three-body kernels}
\label{sec:CoulombKernels}

The two-photon bubble and box diagrams, Figs.~\ref{fig:TwoPhotonCoulomb}(a) and 
\ref{fig:TwoPhotonCoulomb}(b) can be obtained directly from the 
expressions given in Ref.~\cite{Konig:2014ufa} for the analogous topologies 
with a full (screened) Coulomb T-matrix, replacing the latter by the two-photon 
amplitude given in Eq.~\eqref{eq:TC2}.  The kernel functions corresponding to 
two photons attached to the bubble becomes
\begin{equation}
 K_\text{bub}^{(2)}(E;k,p) = {-\frac{\MN}{2\pi^2}}
 \times\frac12\int_{-1}^1\dd\!\cos\theta
 \,\mathcal{I}_\text{bubble}^{(2)}(E;\vk,\vp)
\label{eq:K-bub-2}
\end{equation}
with
\begin{multline}
 \mathcal{I}_\text{bubble}^{(2)}(E;\vk,\vp)
 = \int_0^\infty \dd q\,q^2\,
 T_{C}^{(2)}\!\left(\ii\sqrt{3q^2/4-\MN E};\veck,\vecp\right) \\
 \times\frac1{\sqrt{F}}\left[\log\!\left(\frac{A+B}{A-B}\right)
 - \log\!\left(\frac{B(2C+D)-A(D+2E)+2\sqrt{F}\sqrt{C+E+D}}
 {B(2C-D)-A(D-2E)+2\sqrt{F}\sqrt{C+E-D}}\right)\right] \,,
\label{eq:BubbleInt}
\end{multline}
where
\begin{subequations}\begin{align}
 A &= k^2 + q^2 - \MN E \,, \\
 B &= kq \,, \\
 C &= (p^2 + q^2 - \MN E)^2 - p^2 q^2(1-\cos^2\theta) \, \\
 D &= (p^2 + q^2 -\MN E) \times 2pq\cos\theta \, \\
 E &= p^2 q^2 \,.
\end{align}\label{eq:BubbleInt-ABC}\end{subequations}
and $F = B^2C + A^2E - ABD$.  Likewise, the two-photon box diagram is given by
\begin{equation}
 K_\text{box}^{(2)}(E;k,p) = {-\frac{\MN}{8\pi^3}}
 \times\frac12\int_{-1}^1\dd\!\cos\theta
 \,\mathcal{I}_\text{box}^{(2)}(E;\vk,\vp) \,,
\label{eq:K-box-2}
\end{equation}
where
\begin{equation}
 \mathcal{I}_\text{box}^{(2)}(E;\vk,\vp)
 = \int\dd q\,q^2\,\dd\!\cos\theta'\,\dd\phi'
 \frac{T_{C}^{(2)}\!\left(\ii\sqrt{3q^2/4-\MN E};\veck-\vecq,-\vecp\right)}
 {\left(\vecq^2-\vecq\cdot\veck+\veck^2-\MN E\right)
 \left(\vecq^2-\vecq\cdot\vecp+\vecp^2-\MN E\right)}
\label{eq:I-box-2}
\end{equation}
and the angles are defined via
\begin{subequations}%
\begin{align}
 \vecq\cdot\veck &= qk\cos\theta' \,, \\
 \vecq\cdot\vecp &= pq(\cos\theta\cos\theta'
 +\sin\theta\sin\theta'\cos\phi') \,.
\end{align}
\end{subequations}

A triangle diagram with a full Coulomb T-matrix was not considered in 
Ref.~\cite{Konig:2014ufa}, but it is straightforward to write down 
(\cf~Refs.~\cite{Kok:1979aa,Kok:1981aa,Ando:2010wq}).  The S-wave projected 
result is
\begin{equation}
 K_\text{tri}^{(2,\text{out})}(E;k,p) = {\frac{\MN}{4\pi^2}}
 \times\frac12\int_{-1}^1\dd\!\cos\theta
 \,\mathcal{I}_\text{tri}^{(2)}(E;\vk,\vp) \,,
\label{eq:K-tri-2}
\end{equation}
with the loop integral
\begin{equation}
 \mathcal{I}_\text{tri}^{(2)}(E;\vk,\vp)
 = \int\dd q\,q^2\,\dd\!\cos\theta'
 \frac{T_{C}^{(2)}\!\left(\ii\sqrt{3q^2/4-\MN E};\vecq,\vecp+\veck/2\right)}
 {\left(\MN E -3\vk^2/4-\vq^2\right)
 \left(\vk^2+\vp^2+\vk\cdot\vp-\MN E\right)} \,.
\label{eq:I-tri-2}
\end{equation}
In this case, $\theta'$ is the angle between $\vq$ and $\vecp+\veck/2$.  
The reversed diagram is given by $K_\text{tri}^{(2,\text{in})}(E;k,p) = 
K_\text{tri}^{(2,\text{out})}(E;p,k)$.

Note that no photon-mass regularization is required in the above expressions.  
From Eq.~\eqref{eq:TC2} one finds that
$T_C^{(2)}(k;\vecp,\vecq) \propto x^{{-}1/2} + \OO(x^{1/2})$ with $x = 
(\vecp - \vecq)^2$, which is directly integrable over the angle between $\vecp$ 
and $\vecq$.

\end{document}